\documentclass[pre,reprint,superscriptaddress,titlepage,floatfix,showkeys]{revtex4-2}
\usepackage{amsmath}
\usepackage{bm}
\usepackage{multirow}
\usepackage[latin9]{inputenc}
\usepackage{latexsym, amsfonts, amssymb, wasysym}
\usepackage[dvipsnames]{xcolor}
\usepackage{graphicx}
\usepackage{nicefrac}
\usepackage{amsfonts}
\usepackage[version=3]{mhchem}
\usepackage{subfigure}
\usepackage{multirow}
\usepackage{tabularx}
\usepackage{array}
\usepackage{units}
\usepackage{braket}
\usepackage{bm,times}
\usepackage{booktabs}
\usepackage{enumitem}
\usepackage{mathtools}
\usepackage{epstopdf}
\usepackage{threeparttable}
\usepackage{tensor}
\usepackage[colorlinks = true, linkcolor = blue, urlcolor  = blue, citecolor = blue, anchorcolor = blue]{hyperref}

\def\be{\begin{equation}}
\def\ee{\end{equation}}
\def\bea{\begin{eqnarray}}
\def\eea{\end{eqnarray}}

\def\deg{$^\circ$}

\def\asi{{\it a}-Si}

\def\csi{{\it c}-Si}
\def\asih{{\it a}-Si:H}

\begin{document}

\title{
{\it Ab initio} hydrogen dynamics and the morphology of 
voids in amorphous silicon
}

\author{
Parthapratim Biswas 
}
\email[Corresponding author:\,]{partha.biswas@usm.edu}

\author{
Dil Limbu
} 

\affiliation{
Department of Physics and Astronomy, The University of Southern 
Mississippi, Hattiesburg, Mississippi 39406, USA
} 


\begin{abstract}
This paper presents an {\it ab initio} study of hydrogen 
dynamics inside nanometer-size voids in {\it a}-Si within 
the framework of the density-functional theory for a 
varying hydrogen load of 10 to 30 H atoms/void at the low 
and high temperature of 400 K and 700 K, respectively.  
Using the local density approximation and its 
generalized-gradient counterpart, the dynamics 
of hydrogen atoms inside the voids are examined with an 
emphasis on the diffusion of H atoms/molecules, and 
the resulting nanostructural changes of the void 
surfaces. 
The results from simulations suggest that the microstructure 
of the hydrogen distribution on the void surfaces and the 
morphology of the voids are characterized by the presence 
of a significant number of monohydride Si--H bonds, along 
with a few dihydride Si--H$_2$ configurations.  The study 
also reveals that a considerable number (about 10--45 at.~\%) 
of total H atoms inside voids can appear as H$_2$ molecules 
for a hydrogen load of 10--30 H atoms/void.
The approximate shape of the voids is addressed from a 
knowledge of the positions of the void-surface atoms 
using the convex-hull approximation and the Gaussian 
broadening of the pseudo-atomic surfaces of Si and H atoms. 
\end{abstract} 

\maketitle   % please do not remove

\section{Introduction}

Hydrogen plays an important role in the electronic and 
optical properties of amorphous silicon ({\asi})~\cite{Street2009}. 
While the addition of a small amount (about 6--12 at.~\%) of hydrogen 
in {\asi} is particularly beneficial for producing device-grade 
samples of {\asih}, via passivation of coordinating 
defects (e.g., 3-fold-coordinated Si atoms 
or dangling bonds), the presence of too 
much hydrogen in {\asi} can be detrimental to the 
electronic and optical properties of {\asih}-based 
devices. Thin films of {\asih} are often employed for 
surface passivation of crystalline silicon, which is 
useful for the generation of high open-circuit voltages 
in silicon-heterojunction solar cells~\cite{HJT2012}. 
A relatively high hydrogen content, in void-rich 
environment of {\asih}, is preferred for this 
purpose to achieve high-quality {\asih/\csi} 
interfaces~\cite{Ge2012}, indicating the benevolent 
role of H in {\asi}. By contrast, the structure of 
{\asih} is altered by the light-induced creation of 
metastable defect states, known as the Staebler-Wronski 
effect (SWE)~\cite{Staebler1977}, which adversely affects 
the performance of {\asih}-based solar cells. Nuclear 
magnetic resonance (NMR) studies~\cite{Su2002} on {\asih} samples 
have indicated that the motion of H atoms, which are produced 
by the light-induced breaking of Si-H bonds, plays an 
important part in recovering the photovoltaic stability 
of the affected {\asih} samples upon annealing at 
350{\deg}--400{\deg} C~\cite{Bobela2011}. Thus, an 
understanding of the motion of H atoms in {\asi} in 
the presence of inhomogeneities is of crucial 
importance to address the photovoltaic stability 
of {\asih} upon light irradiation. 

While a number of earlier theoretical/computational 
studies on the microstructure of hydrogen 
distributions in {\asih}~\cite{Biswas2007,Chakraborty2009,Biswas2011,Biswas2017,Biswas2020,Limbu2020} have focused on the silicon-hydrogen 
bonding configurations in the bulk network environment 
of {\asi} and their effects on structural, electronic, 
and optical properties of the material, there exist only a few 
computational studies that directly address the role 
of non-bonded hydrogen (NBH) and voids on the network 
structure of {\asih}~\cite{Biswas2017,BiswasJAP2014,Guerrero2020}. 
Sekimoto et al.~\cite{Sekimoto2016} 
have recently shown experimentally that the presence 
of large amount of non-bonded hydrogen (e.g., H$_2$ 
molecules inside voids) can broaden the vacancy-size 
distribution and enhance the size of the optical 
gap~\cite{Sekimoto2018} in {\asi}. The density of {\asih} 
has been also found to be somewhat dependent on the 
number of NBH, and hence on the distribution of hydrogen 
inside nanometer-size voids. Following our recent 
studies on the temperature-induced nanostructural 
evolution of voids in {\asi}~\cite{Biswas2020}, and 
its effect on the intensity of small-angle X-ray 
scattering (SAXS)~\cite{Paudel2018}, the present 
study focuses on accurate calculations of the 
atomistic dynamics of hydrogen inside voids 
obtained from the density-functional theory.  
The emphasis here is on the dynamical aspects of hydrogen 
motion inside voids on the time scale of several 
picoseconds at low and high temperature in the 
environment of a varying concentration of H atoms. 
The formation and dissociation of Si--H bonds on 
the surface of voids are discussed from a 
kinetic point of view. The movement of H atoms 
inside nanometer-size voids and its resulting 
effects on the hydrogen microstructure and 
the reconstruction of the void surfaces are 
also addressed in this study. 

The presence of voids in the amorphous matrix suggests 
that, for an accurate determination of the motion of 
H atoms within a void, one must take into account the 
inhomogeneities in the electronic charge distribution 
in the vicinity of void surfaces. The standard protocol in 
density-functional calculations advises us to address 
the problem by computing the self-consistent-field (SCF) 
solution of the Kohn-Sham (KS) equation, and the 
presence of voids suggests that 
the generalized-gradient approximation (GGA) should be 
employed in order to deal with the atomic density fluctuations 
near the void surfaces. We shall therefore address the 
problem using the GGA and compare the results with 
those from the local density approximation (LDA).  The 
presence of weak dispersion forces, which are often 
included via the Van der Waals corrections, are not 
taken into account in this study owing to the computational 
complexity of the problem and the somewhat limited 
accuracy of local basis functions that we have used in 
this study. 

The rest of the paper is organized as follows. Section II 
provides a short description of the method for generating
{\asi} models with hydrogenated voids, using the
Wooten-Winer-Weaire (WWW) method and {\it ab initio}
molecular dynamics simulations (AIMD). The results are
discussed in sec.\,III, with an emphasis on the dynamics
of H atoms inside voids in IIIA, kinetics of Si--H 
bond formation and dissociation in IIIB, and the 
microstructure of the hydrogen distribution for a 
varying number of hydrogen atoms in IIIC. A brief 
discussion on the shape of the voids is also provided 
in IIIC.  This is followed by the conclusions of the 
study in sec.\,IV.

\section{Method} 
For AIMD simulations of hydrogen dynamics inside voids in 
{\asi}, we started with four independent 1000-atom 
models of {\asi}, obtained from the WWW method~\cite{W3,Barkema2000}. 
A spherical void of radius 5 {\AA} was created at the 
center of each model and the geometry of the resulting 
structure was thoroughly relaxed, using the first-principles 
density-functional code {\sc Siesta}~\cite{Siesta2002}. A 
number of models with a hydrogenated void were then produced 
by adding 10, 20, and 30 H atoms inside the central cavity 
so that the mass density of the final models is about 2.22 
g.cm$^{-3}$. A single void of radius 5 {\AA} corresponds 
to a number density of $4.82\times 10^{19}$ voids.cm$^{-3}$ 
and a void-volume fraction of 2.52 at.\,\%, which are close to 
the values reported in experiments~\cite{Mahan2001,Young2007}.  
The H atoms were initially distributed in such a way that 
they were at a distance of at least 2 {\AA} from Si atoms 
and 1 {\AA} from each other. The silicon atoms within the 
spherical region of radius between 5 {\AA} and 8 {\AA} from 
the center of the voids are labeled as the void-surface 
atoms for the analysis of the void surface upon annealing 
and total-energy optimization of the systems~\cite{void}. 

To study hydrogen diffusion and the microstructure of hydrogen 
distribution on the void surfaces, AIMD simulations were 
performed at 400 K and 700 K to examine the temperature 
dependence of H dynamics and the resulting effects on 
the morphological structure of the void surfaces, using 
the density-functional code {\sc Siesta}. 
{\sc Siesta} employs local basis functions, based on numerical 
pseudoatomic orbitals~\cite{Artacho2001}, and the norm-conserving 
Troullier-Martins pseudopotentials~\cite{TM1991} in order 
to obtain the self-consistent-field (SCF) solution of the Kohn-Sham 
equation in the density-functional theory. Electronic and 
exchange correlations between electrons were taken into 
account via the LDA and the GGA, by using the Perdew-Zunger~\cite{LDA} 
and the PBE~\cite{PBE} parameterizations for the LDA and 
the GGA, respectively.  
Owing to the computational complexity
of the problem involving a large number of atoms and the
necessity for simulating a reasonably long-time dynamics
for studying the equilibrium distribution of H atoms, 
using the SCF solution of the KS equation, the AIMD 
simulations were conducted using the single-zeta (SZ) basis 
functions for Si atoms and the double-zeta-polarized 
(DZP) basis functions for H atoms.  The high computational 
cost associated with the calculation of total energy 
and forces restricts us to use SZ basis functions for 
Si atoms during the course of AIMD simulations. The 
subsequent total-energy optimizations, however, were 
conducted using the double-zeta (DZ) basis functions 
for Si atoms until the total force on each atom was 
less than or equal to $5\times10^{-3}$ eV.{\AA$^{-1}$}. 
During AIMD simulations, the temperature of the 
system was controlled in NVT ensembles by using 
the Nos\'e thermostat~\cite{Nose1984}, and a time 
step of 0.8 fs was used to ensure that the movement 
of light H atoms can be described accurately at
the high temperature of 700 K.
The evolution of the
system was monitored and recorded for a total time of
20 ps. The simulation procedure was repeated for each
of the four independent models with a void at the 
center, indicated by M1 to M4 hereafter, for a 
hydrogen load of 10, 20, and 30 H atoms/void. 
In each case, we have used a different
random distribution of H atoms within the void in order
to gather as much as statistics as possible. The final
results were obtained by averaging over four 
independent configurations.

\begin{figure}[t!]
\centering
\includegraphics[width=0.4\textwidth]{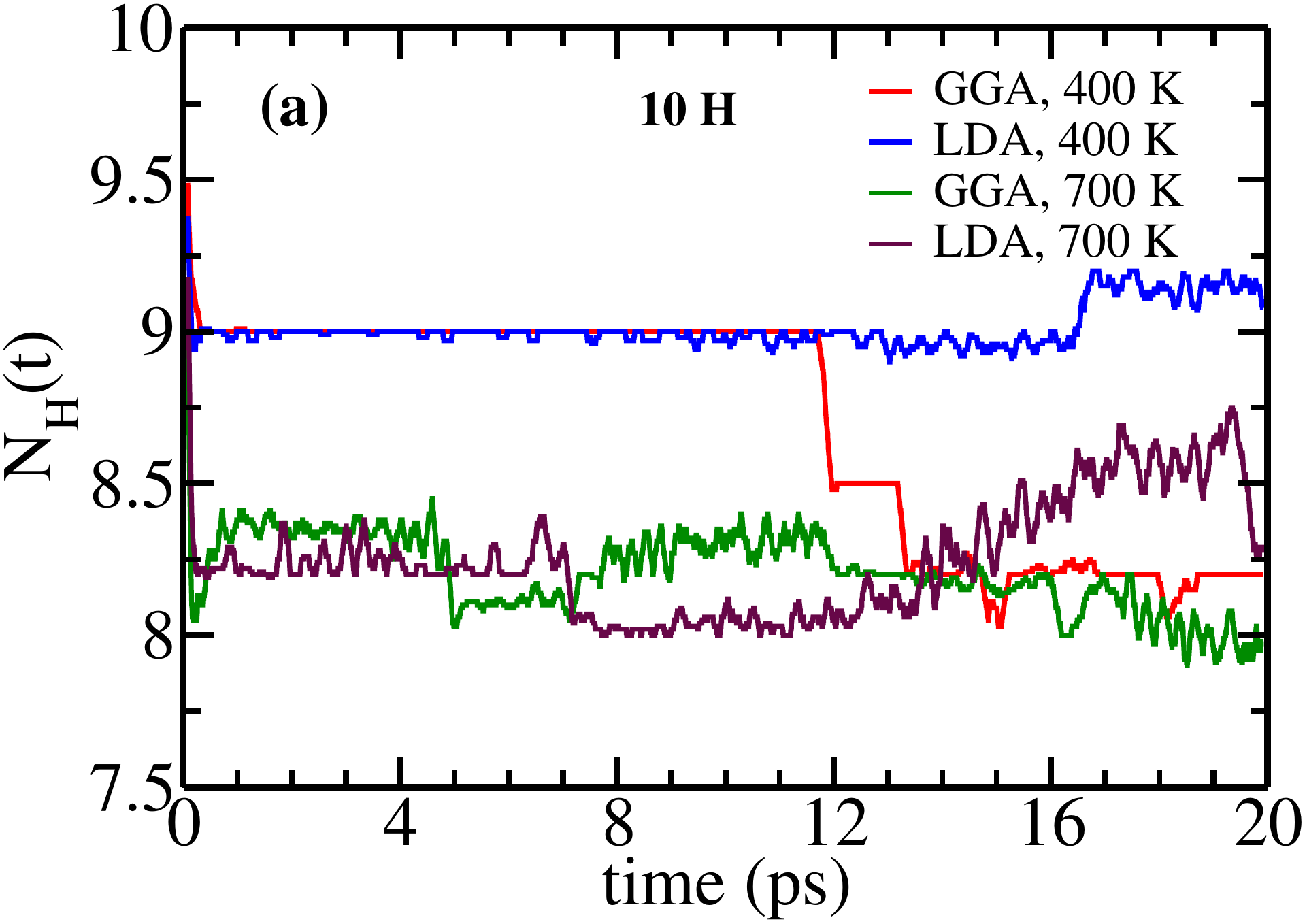}
\includegraphics[width=0.4\textwidth]{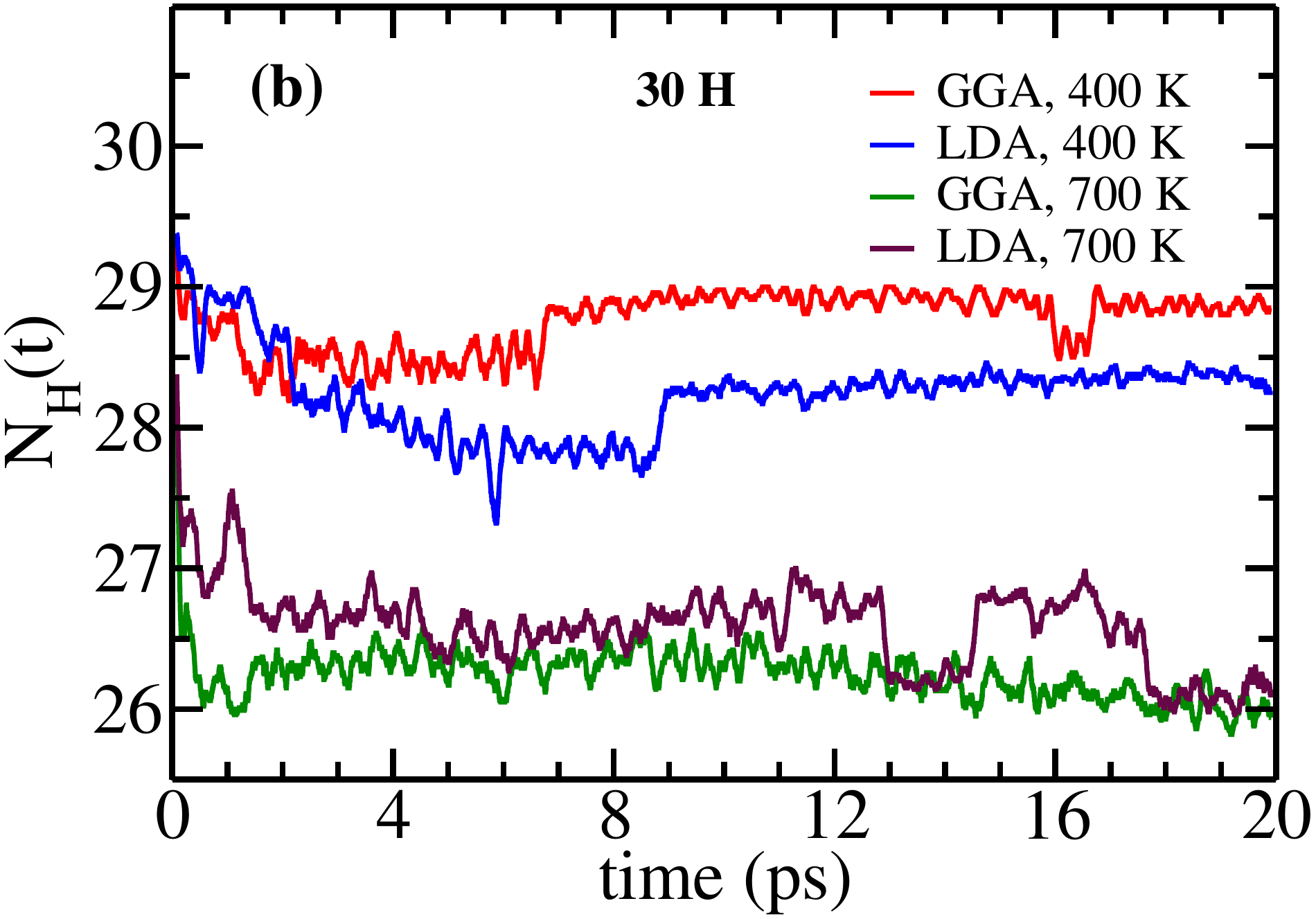}
\caption{\label{FIG1}
The time evolution of the average number of H atoms 
(including H$_2$ molecules) within a void at 400 K 
and 700 K. The results correspond to a hydrogen load 
of: (a) 10 H atoms/void; (b) 30 H atoms/void. The 
average is taken over 4 voids for each temperature 
from independent simulations. 
}
\end{figure}

\section{Results and Discussion}

\subsection{Hydrogen dynamics inside voids in {\asi}}
In this section, we discuss the results from the density-functional 
calculations for studying the dynamics of H atoms/molecules 
inside voids in {\asi}. To this end, we mostly focus 
on those H atoms/molecules which are within the cavity during 
the entire course of simulations at a low and high 
temperature of 400 K and 700 K, respectively. Figure~\ref{FIG1} 
shows the time evolution of the average number 
of H atoms (molecules included) within a void-surface 
region of radius 8 {\AA} at 400 K and 700 K. The results 
correspond to two 
different values of the hydrogen load, 10 and 30 H 
atoms/void, and are obtained by averaging over 4 independent 
voids/models using the LDA and GGA. At 400 K, the great 
majority of H atoms stayed within the void, with the 
exception of one or two H atoms that left the void-surface 
region of radius 8 {\AA}. The LDA and GGA dynamics exhibit 
a more or less similar behavior at a given temperature, 
although the GGA tends to knock out one or two more H atoms 
outside the cavity for high H loads at 700 K (see 
Fig.~\ref{FIG1}b). Despite limited statistics, it would 
not be inappropriate to conclude that, 
on average, the LDA and GGA do not differentiate much as 
far as the (average) number of H atoms leaving the voids 
at 400 K is concerned. 

\begin{figure}[t!]
\centering
\includegraphics[width=0.4\textwidth]{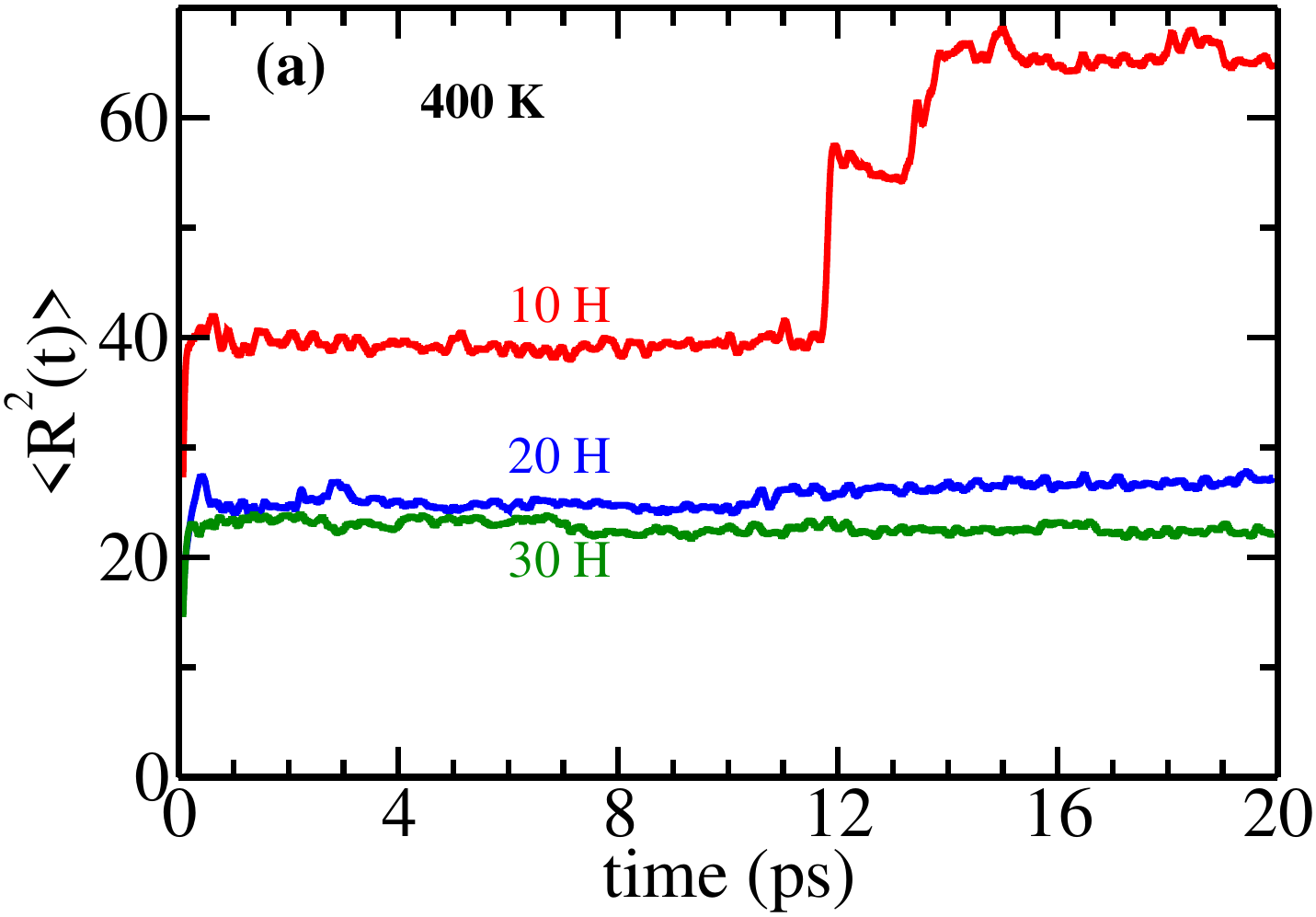}
\includegraphics[width=0.4\textwidth]{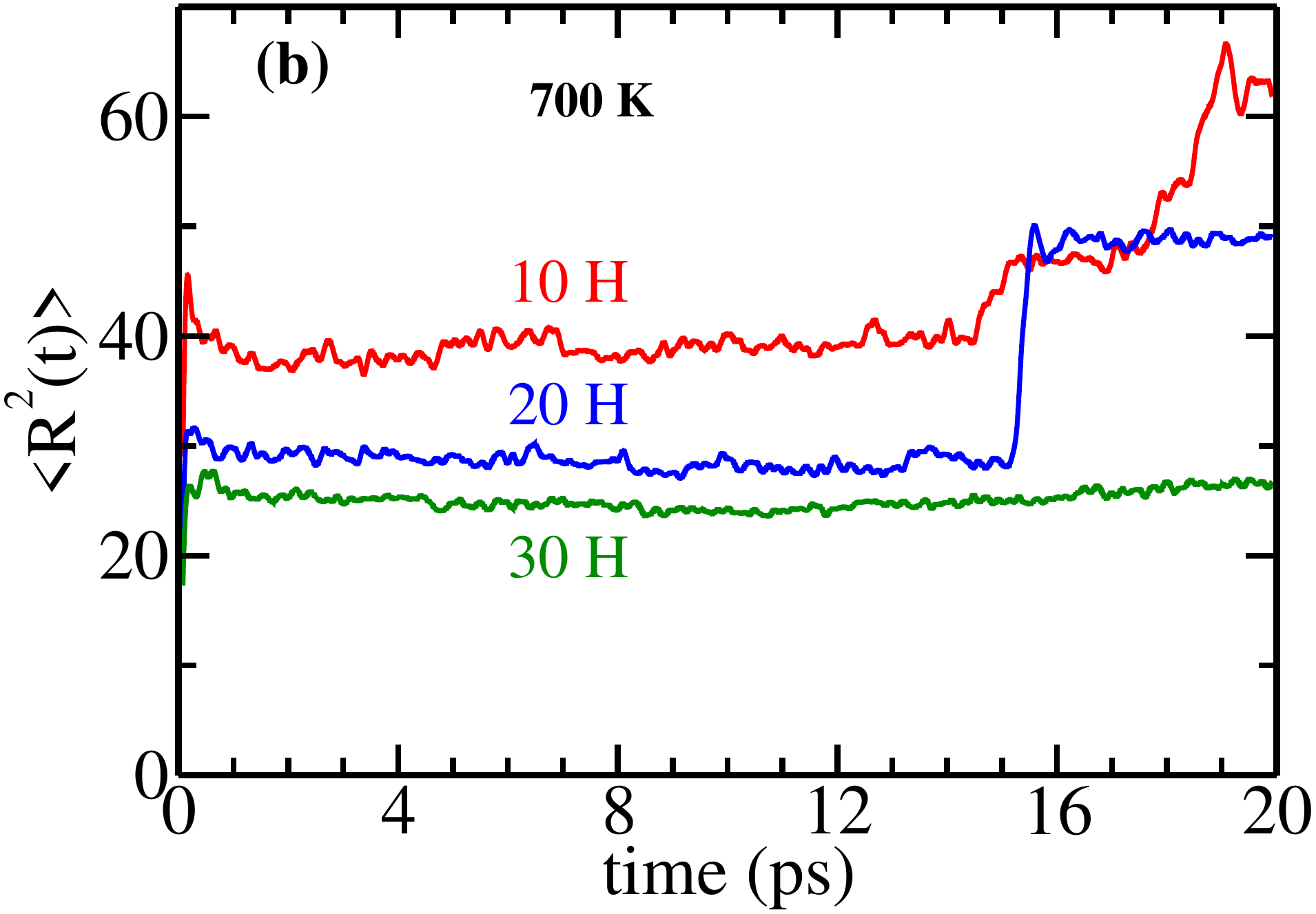}
\caption{
\label{FIG2}
The variation of the mean-square displacement 
(MSD) of H atoms, averaged over {\em all 
hydrogen} in four models, with time for three 
different hydrogen loads as indicated: (a) 
400 K; (b) 700 K.  The results were obtained 
from the self-consistent-field calculations 
with the GGA, and averaged over 4 independent 
voids for each H load and temperature. 
}
\end{figure} 

To study the effect of the hydrogen load on the 
mean-square displacement (MSD) of H atoms at the 
low and high temperature of 400 K and 700 K, 
respectively, we have examined the variation of the MSD with time 
for 10, 20, and 30 H atoms/void using the GGA. 
Once again, the results were 
obtained by averaging over 4 independent voids and 
are presented in Fig.~\ref{FIG2}. The results (in 
Fig.~\ref{FIG2}) lead to the following observations. 
Firstly, the MSD of H atoms decreases with an increasing 
presence H atoms within the voids. This is particularly 
so as the hydrogen load increases from 10 H atoms/void 
to 30 H atoms/void. This reduction in the MSD is 
due to the decrease of the average distance between any 
two H atoms within the cavity for an increasing presence 
of hydrogen within the same volume. Secondly, the sharp 
rise of the MSD for a hydrogen load of 10 H atoms/void 
near 12 ps at 400 K can be attributed to a hydrogen 
atom leaving the void. For a hydrogen load of 10 H 
atoms/void, H atoms can move somewhat more 
freely than those with a load of 20 or 30 H atoms/void. 
This makes it possible for few H atoms to diffuse 
rapidly in the vicinity of the void surface, and 
eventually to escape the void region -- and the 
simulation cell in one or two cases -- depending 
upon the temperature of the system. In contrast, the 
presence of too many H atoms for a high value of H load 
can impede the diffusion process and thus reduces the 
MSD at a given temperature. Thirdly, the motion of 
H atoms are affected by the temperature of the system as well.
This is evident in Fig.~\ref{FIG2}(b), where 
the MSD for a load of 20 H atoms/void was 
observed to increase at 700 K during 15 to 
16 ps and it continued to remain so until the end of 
the simulation at 20 ps. This observation contrasts 
with the case of 30 H atoms/void at the same 
temperature, where none of the H atoms was found 
to leave the void but remained within a root-mean-square 
(RMS) distance of 5 {\AA} from the center of 
the void. 
Finally, one may note that the 
available statistics do not permit us to comment 
on the dynamical behavior of few H atoms that 
diffuse out of the cavity at an earlier time 
(e.g., the H atom in Fig.~\ref{FIG2}(a) for a hydrogen 
load of 10 atoms/void), based on the results from 
4 configurations. 
The movement of H atoms is driven by a combination 
of factors, such as the concentration of H atoms 
in a void, temperature, and the disorder in the atomic 
distribution on the void surfaces.  
Since we are not particularly interested in the 
dynamical behavior of few H atoms that escape 
the void-surface region of 8 {\AA}, we will 
not discuss this further. 

\begin{figure}[t!]
\centering
\includegraphics[width=0.4\textwidth]{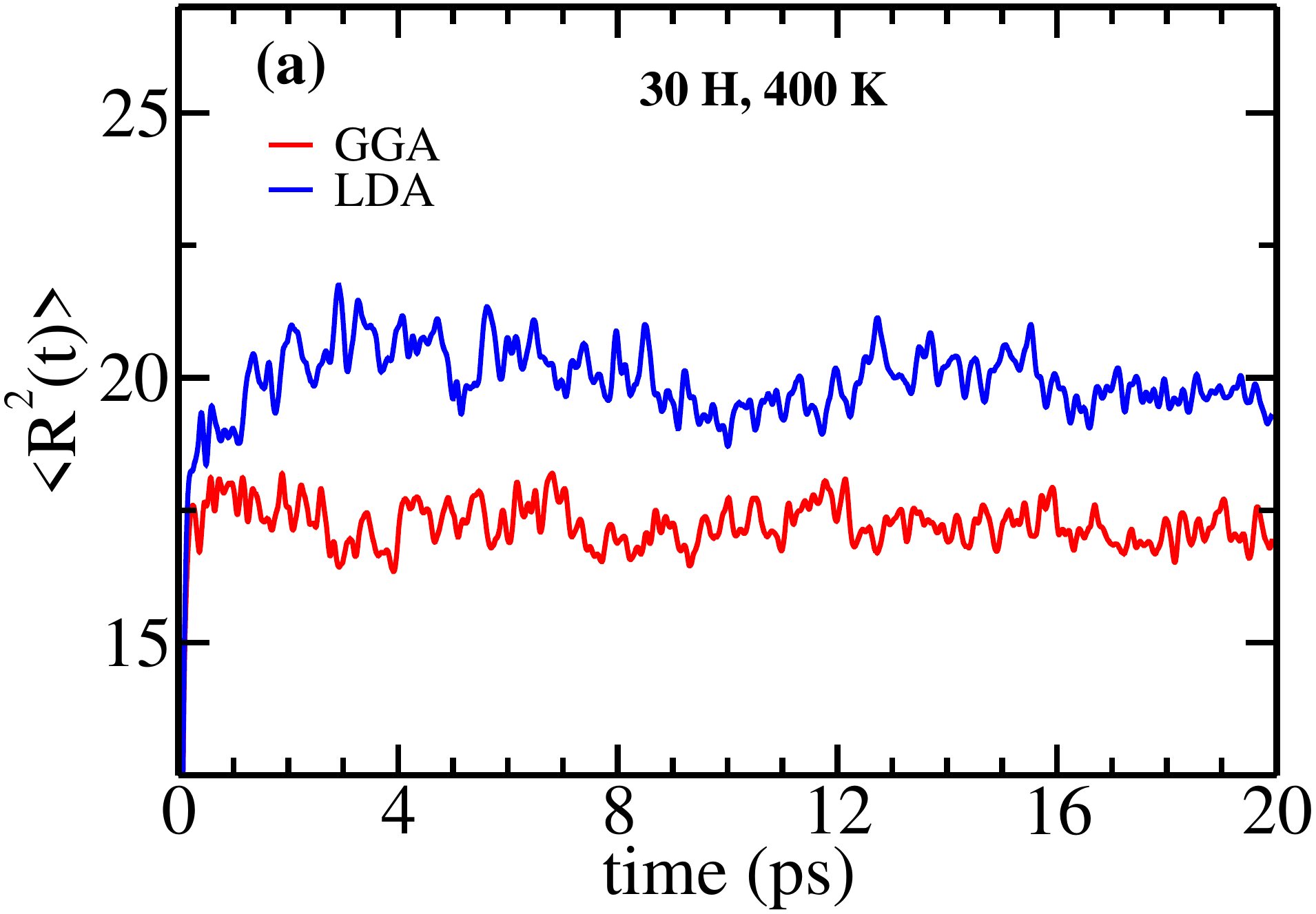}
\includegraphics[width=0.4\textwidth]{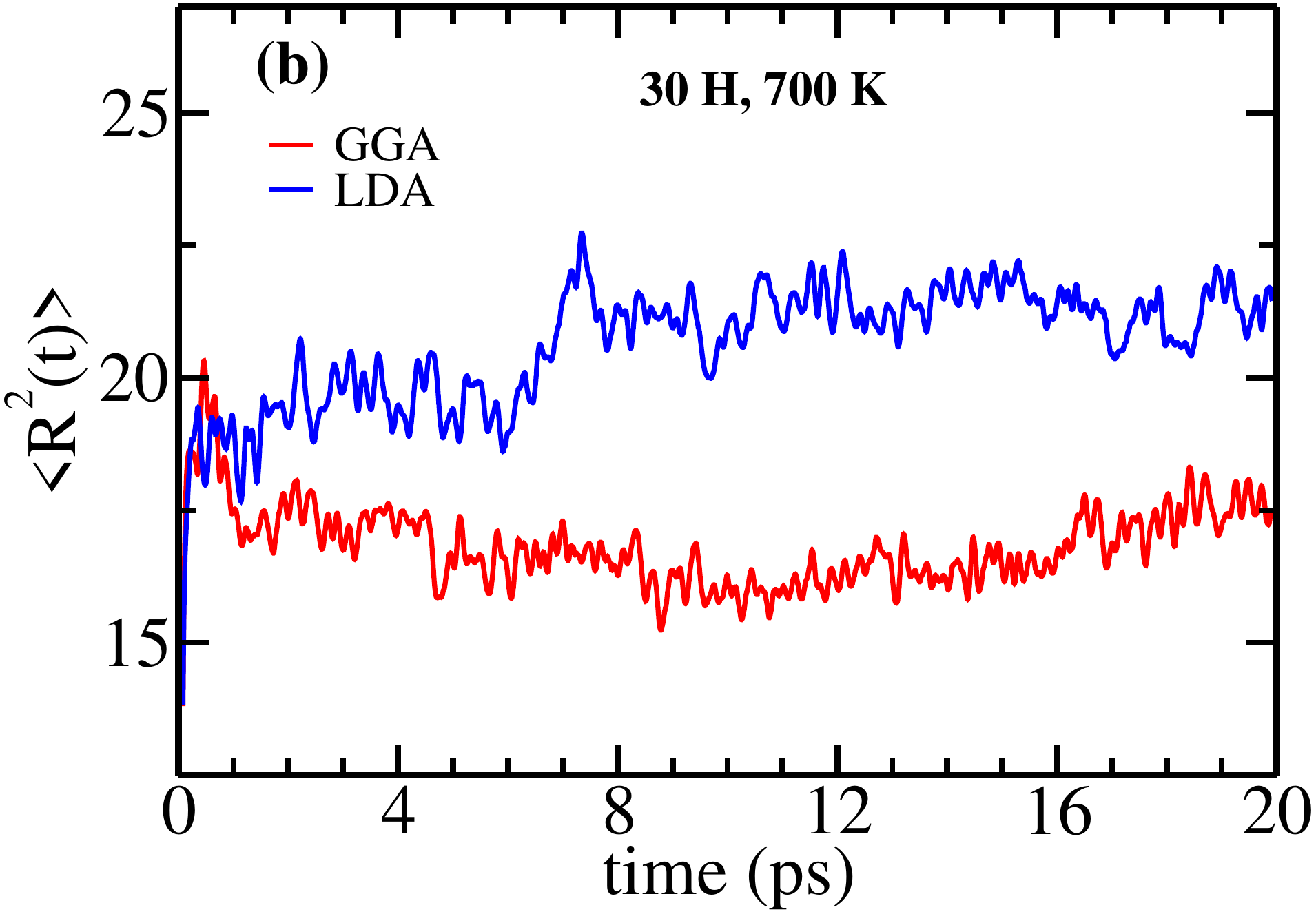}
\caption{
\label{FIG3}
The time evolution of the MSD of H atoms {\em within} 
voids for a hydrogen load of 30 H atoms/void in the 
LDA (blue) and the GGA (red). The results for 400 K 
and 700 K are shown in (a) and (b), respectively. 
}
\end{figure}

\begin{figure}[t!]
\centering
\includegraphics[width=0.4\textwidth]{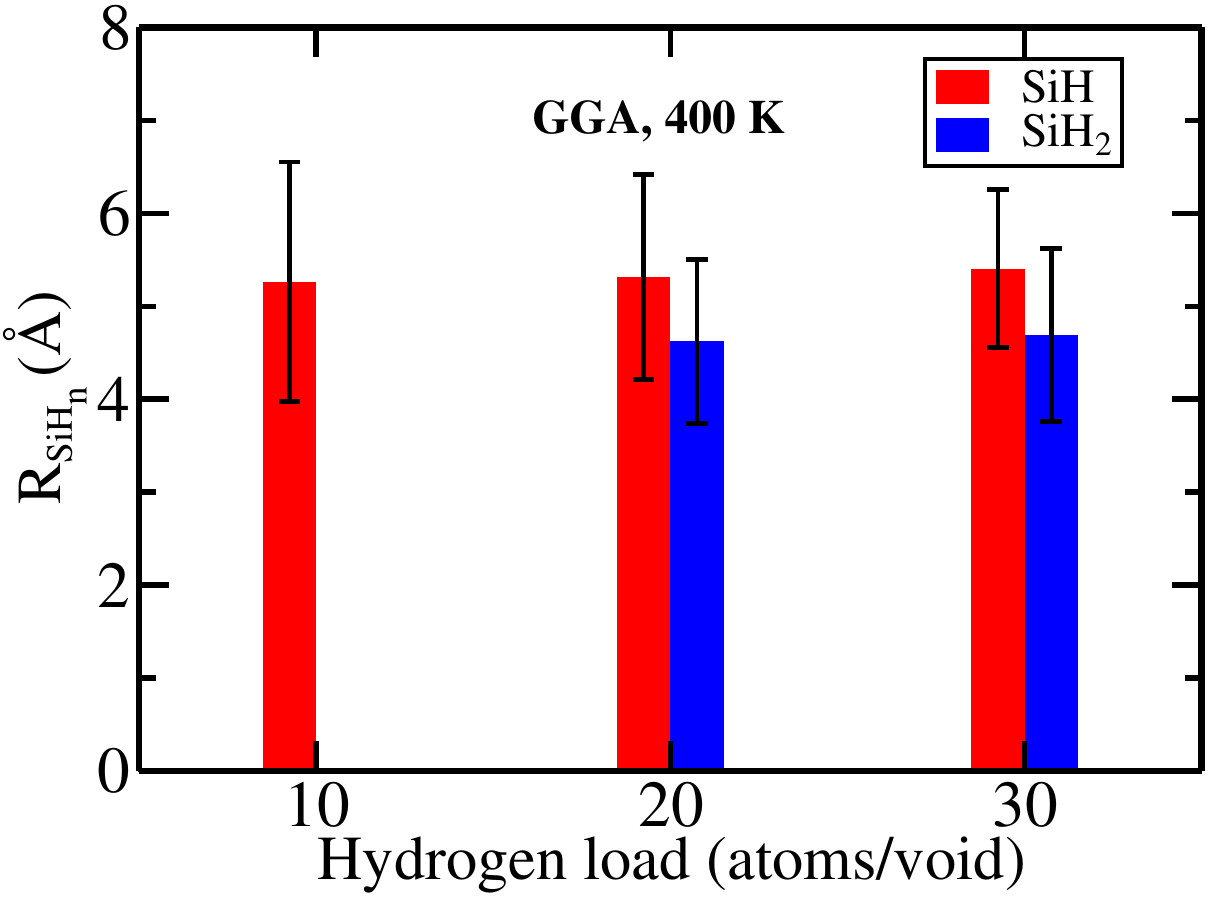}
\caption{\label{FIG4}
The average (radial) distances of SiH and 
SiH$_2$ bonds on the void surface from the 
center of the void. The dihydride SiH$_2$ 
bonds have been found to form on the walls 
of the voids at a distance which is closer 
to the center of the void than their SiH 
counterpart.  
}
\end{figure}

Having discussed the variation of the MSD with respect to the 
hydrogen load, we now address to what extent the 
exchange-correlation (XC) approximation may affect the motion 
of H atoms inside the voids as far as the MSD is 
concerned. Figure \ref{FIG3} shows the evolution of 
the MSD with time for a hydrogen load of 30 H atoms/void 
at 400 K and 700 K. 
Since we are interested in the dynamics of H atoms inside 
the void, and the resulting hydrogen microstructure of 
the void surface, the MSD was computed using only those 
H atoms that were {\em inside} the void region of radius 
8 {\AA} for the entire duration of simulation. 
The plots in Fig.\,\ref{FIG3} suggest that the LDA slightly 
overestimates the MSD (of H atoms) compared to the value 
obtained from the generalized-gradient approximation (GGA). 
The RMS values of the displacement of H atoms inside the 
voids in the LDA and the GGA at 400 K have been 
found to be about 4.5 {\AA} and 4.15 {\AA}, respectively.  A more or less 
similar observation applies to the high-temperature 
dynamics at 700 K, as shown in Fig.\,\ref{FIG3}(b).  It thus appears 
that the MSD of the H atoms within the void is not particularly 
strongly affected by the XC approximation at 400 K and 700 K.  
This is due to the fact that the majority of 
H atoms within a void reside as bonded 
hydrogen (to Si atoms) on the void surface, 
mostly in the form of Si--H configurations, 
and few Si--H$_2$ configurations, along with 
a few H$_2$ molecules within the cavity. Since the calculation 
of the MSD excludes a few mobile H atoms that have 
already left the void region, the remaining bonded 
H atoms (on the void-surface) show a more or less 
converged value of the MSD during the course of 
simulations. Thus, in a sense, a more or less 
converged value of the MSD (in Fig.\,\ref{FIG3}) is 
reflective of the distribution of bonded H atoms on the 
walls of the void. We shall see later that 
a slightly larger value of the MSD of H atoms that 
we observe here for the LDA case (see Fig.~\ref{FIG3}) 
results from a minor expansion of the void surface 
in the LDA calculations. Following Sekimoto et 
al.~\cite{Sekimoto2016}, one may conclude that 
the expansion originates from the stress on the 
void surface due to the presence of a significant 
number of Si--H bonds in the LDA calculations.  
Finally, it may be mentioned that the dihydride 
Si--H$_2$ configurations are found on 
the void surface at a distance, which is closer to 
the center of the void than their monohydride 
(Si--H) counterpart. Figure \ref{FIG4} shows the 
average distances of the Si--H$_2$ and Si--H 
bonding configurations, which 
are located on the walls of the voids, from the 
center of the voids for the GGA at 400 K.  
This observation is consistent with the results 
reported by Kageyama et al.\,\cite{Kageyama2011} 
from dielectric measurements. 

\begin{figure}[t!]
\centering
\includegraphics[width=0.4\textwidth]{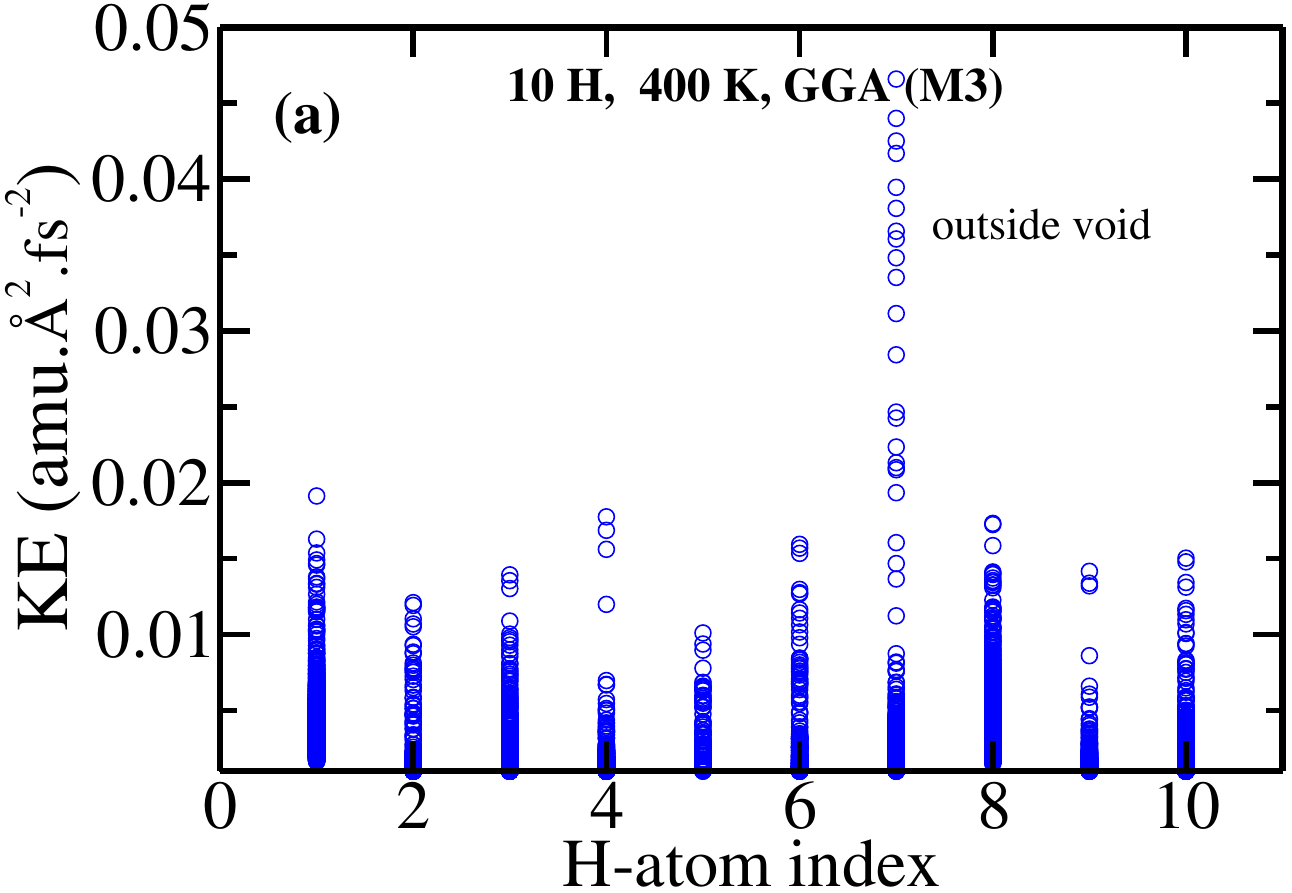}
\includegraphics[width=0.4\textwidth]{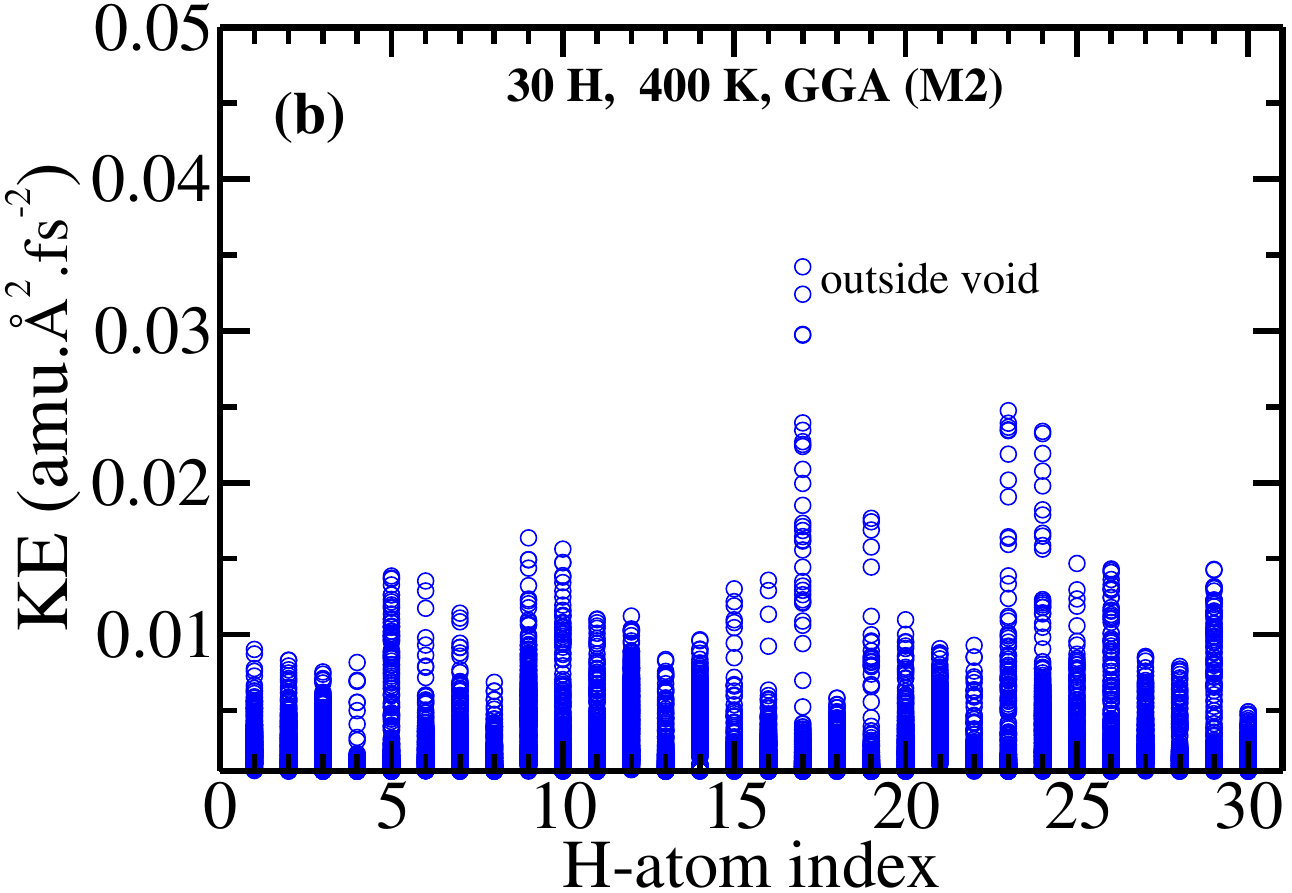}
\caption{\label{FIG5}
The instantaneous kinetic energy (KE) of H atoms in 
two voids (in models M2 and M3) at 400 K, obtained 
from the GGA. The results for the void with a 
hydrogen load of (a) 10 H atoms/void, and (b)30 H 
atoms/voids. The atoms with the largest KE, 
H7 (upper panel) and H17 (lower panel), are 
found to diffuse out of the void region.  The 
range and the frequency of KE values are indicated 
by the numbers and shading (of blue circles) along 
the $y$ axis, respectively.}
\end{figure}

\begin{figure}[th!]
\centering
\includegraphics[width=.4\textwidth]{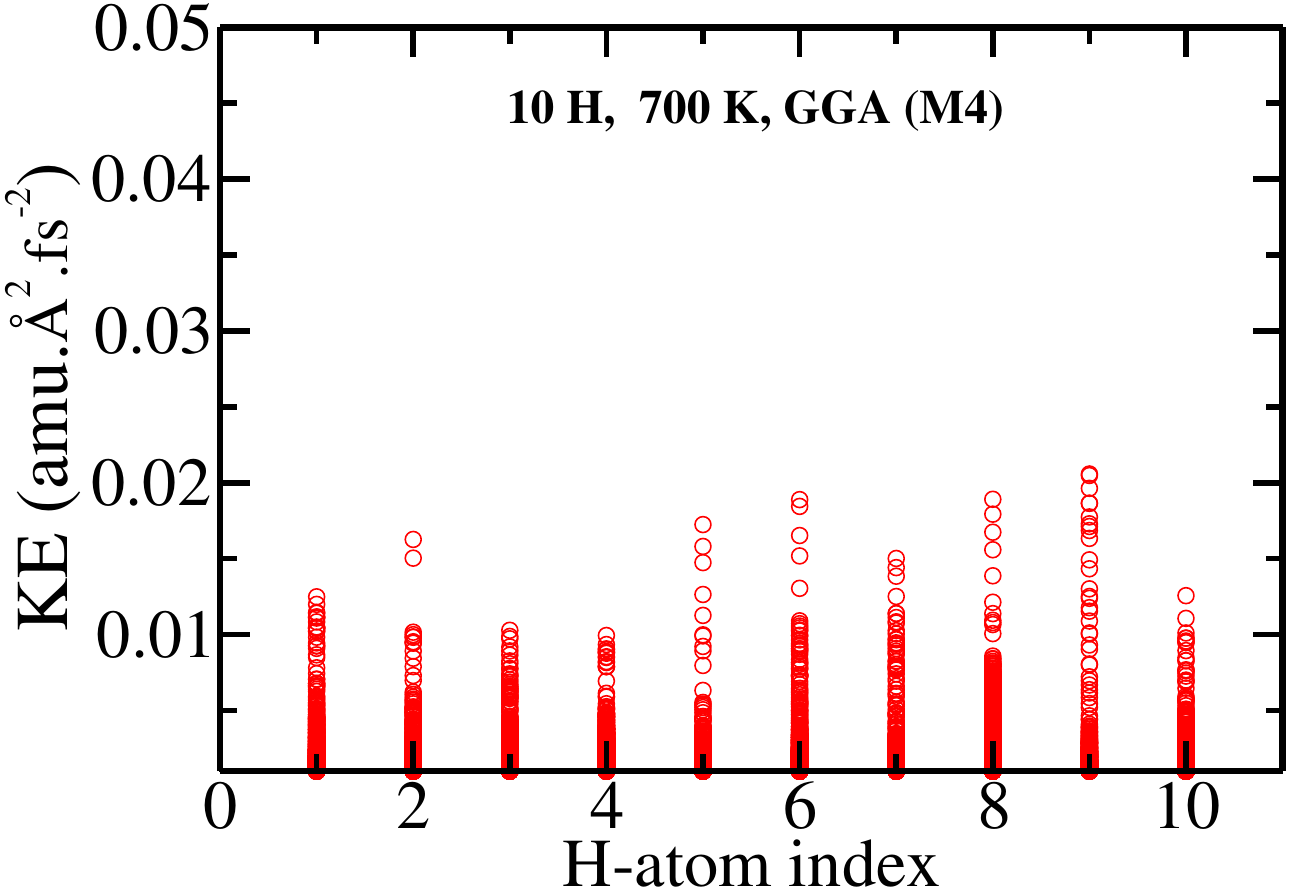}
\caption{
\label{FIG6}
The distribution of kinetic energy (KE) along the $y$ axis for 
H atoms inside a void (in model M4) at 700 K. 
%The relatively low KE values of the H atoms keep all the 
%atoms inside the void during the course of simulation. 
The range and the frequency of KE values are indicated 
by the numbers and shading (of red circles) along the 
$y$ axis, respectively.
} 
\end{figure}

We now briefly discuss the diffusion of a few highly 
mobile H atoms. Earlier in this section, we have seen 
that the MSD of H atoms can increase occasionally 
very rapidly for a hydrogen load of 10 and 20 H 
atoms/void. This behavior of the MSD has been attributed to the 
movement of few H atoms out of the void region, defined 
by a sphere of radius 8 {\AA}. It has been observed that 
such a steep rise of the MSD (e.g., see Fig.\,\ref{FIG2}(a) 
at 12 ps) originates from the high mobility of few H 
atoms inside the hydrogenated voids.  Figure \ref{FIG5} 
shows the plot of the instantaneous kinetic energy (KE) 
of H atoms at 400 K, obtained from using the GGA for 
a period 20 ps. The distribution of the KE values along 
the $y$ axis for a given H atom indicates the range and 
the frequency of the kinetic energy of the atom during 
the course of AIMD simulations.  An analysis of the 
real-space trajectory of the H atoms in the vicinity 
of the void region in models M2 and M3, and the results 
from Fig.~\ref{FIG5} revealed that a few highly mobile 
H atoms, such as H7 in Fig.~\ref{FIG5}(a) and H17 in 
Fig.~\ref{FIG5}(b), left the void region sometime during 
the course of simulations. This observation was found 
to be true for other voids as well, where H atoms were 
observed to leave the void region due to their high 
kinetic energies. Conversely, we have noticed that all 
the H atoms in model M4 remained inside the void throughout 
the course of simulation even at the high temperature 
of 700 K. The results obtained from M4 are shown in 
Fig.~\ref{FIG6}, where the KE values of the H atoms 
are found to be considerably lower than those presented 
in Fig.~\ref{FIG5}.

\begin{figure}[t!]
\centering
\includegraphics[width=0.325\textwidth]{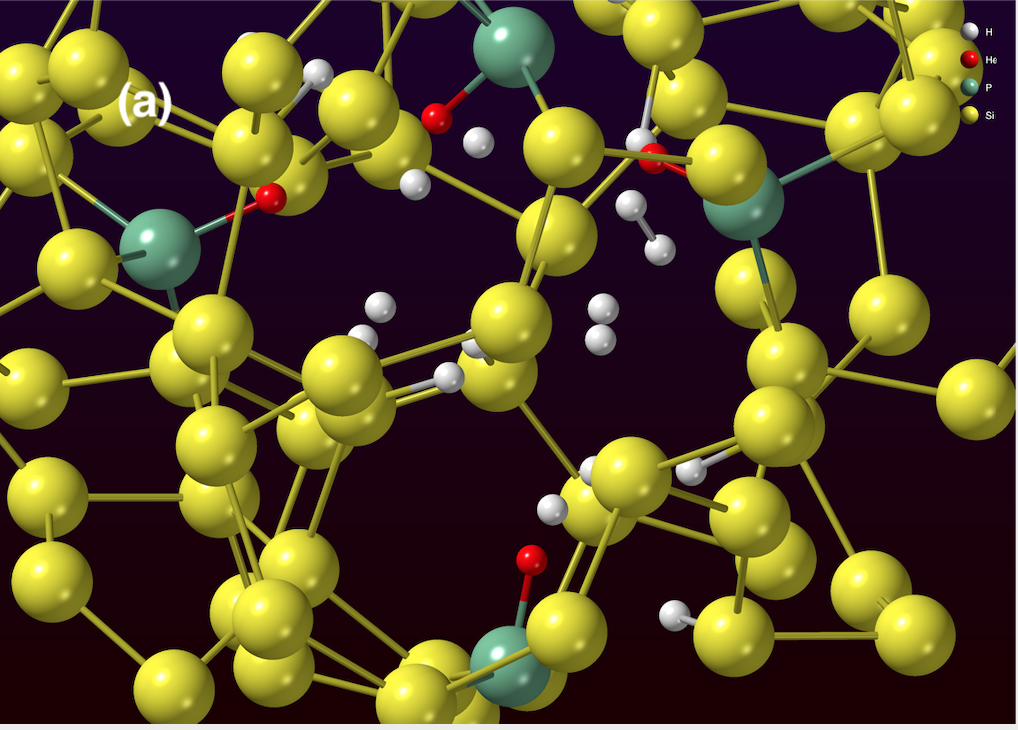}
\includegraphics[width=0.325\textwidth]{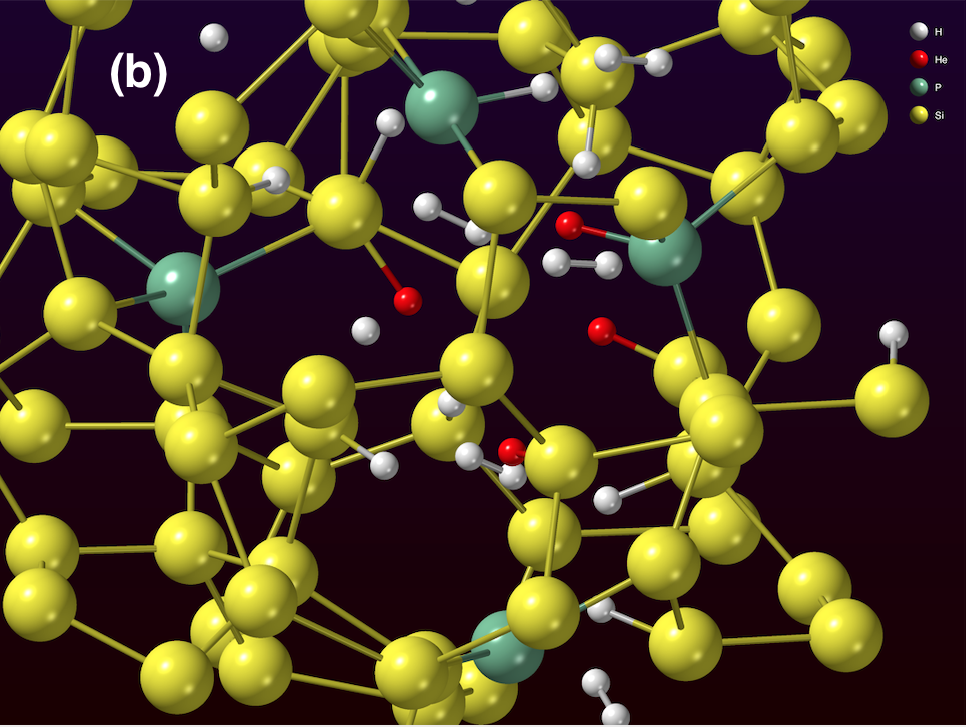}
\caption{
The kinetics of Si--H bond formation and dissociation at 700 K 
for a hydrogen load of 30 H atoms/void obtained from the GGA 
in model M4. (a) The Si--H bonds that dissociated during 
AIMD simulations are shown as green-red pairs at 3 ps. 
(b) The H atoms (red) resulted from the dissociation of SiH 
bonds diffused through the void-surface region to form new 
Si--H bonds, which are indicated as yellow-red pairs at 
18 ps. The presence of a green-red pair suggests that the 
Si--H bond broke and formed again, but with another 
(dissociated) H atom. 
}
\label{FIG7}
\end{figure} 

\begin{figure}[t!]
\centering
\includegraphics[width=0.35\textwidth]{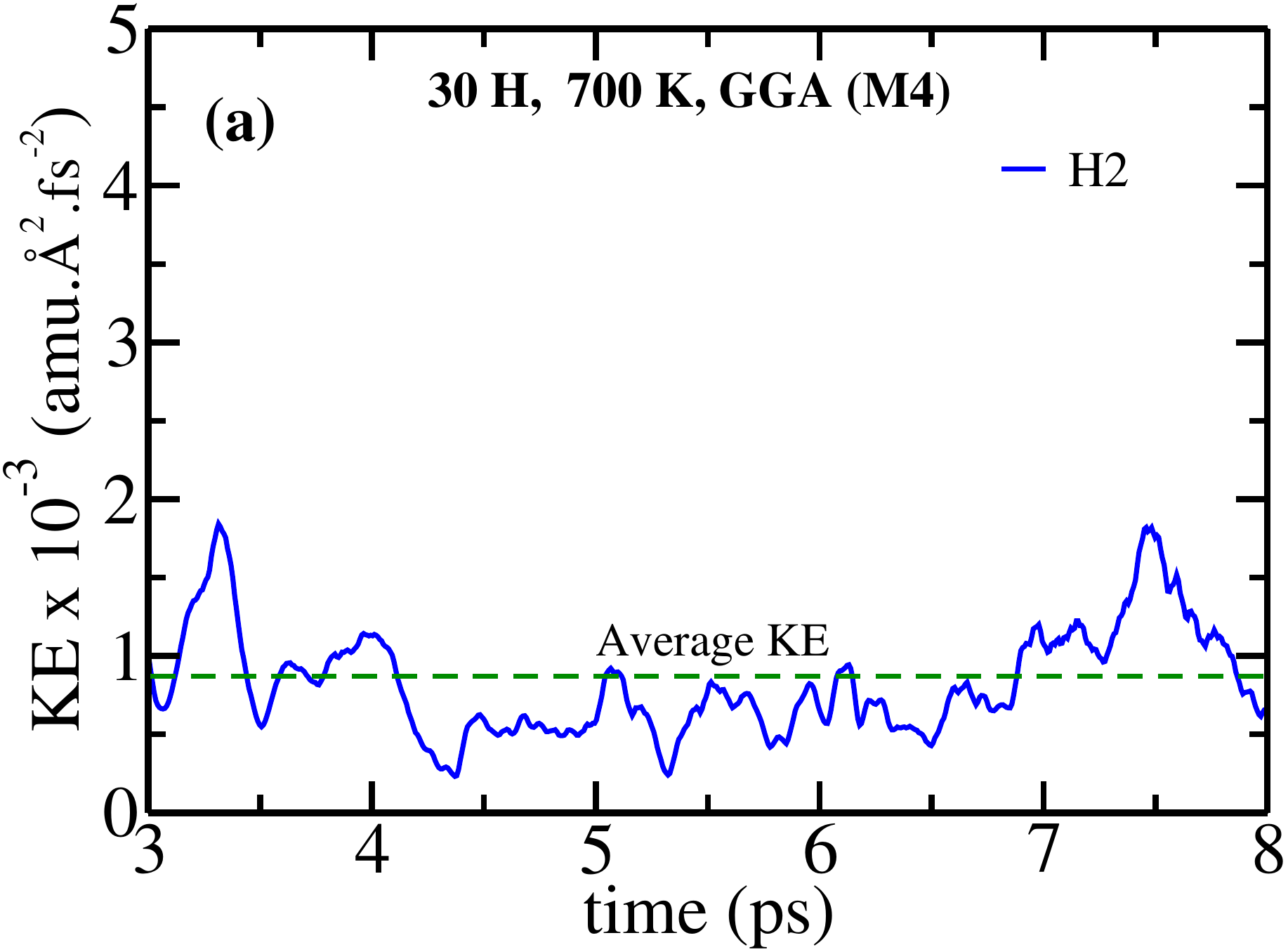}
\includegraphics[width=0.35\textwidth]{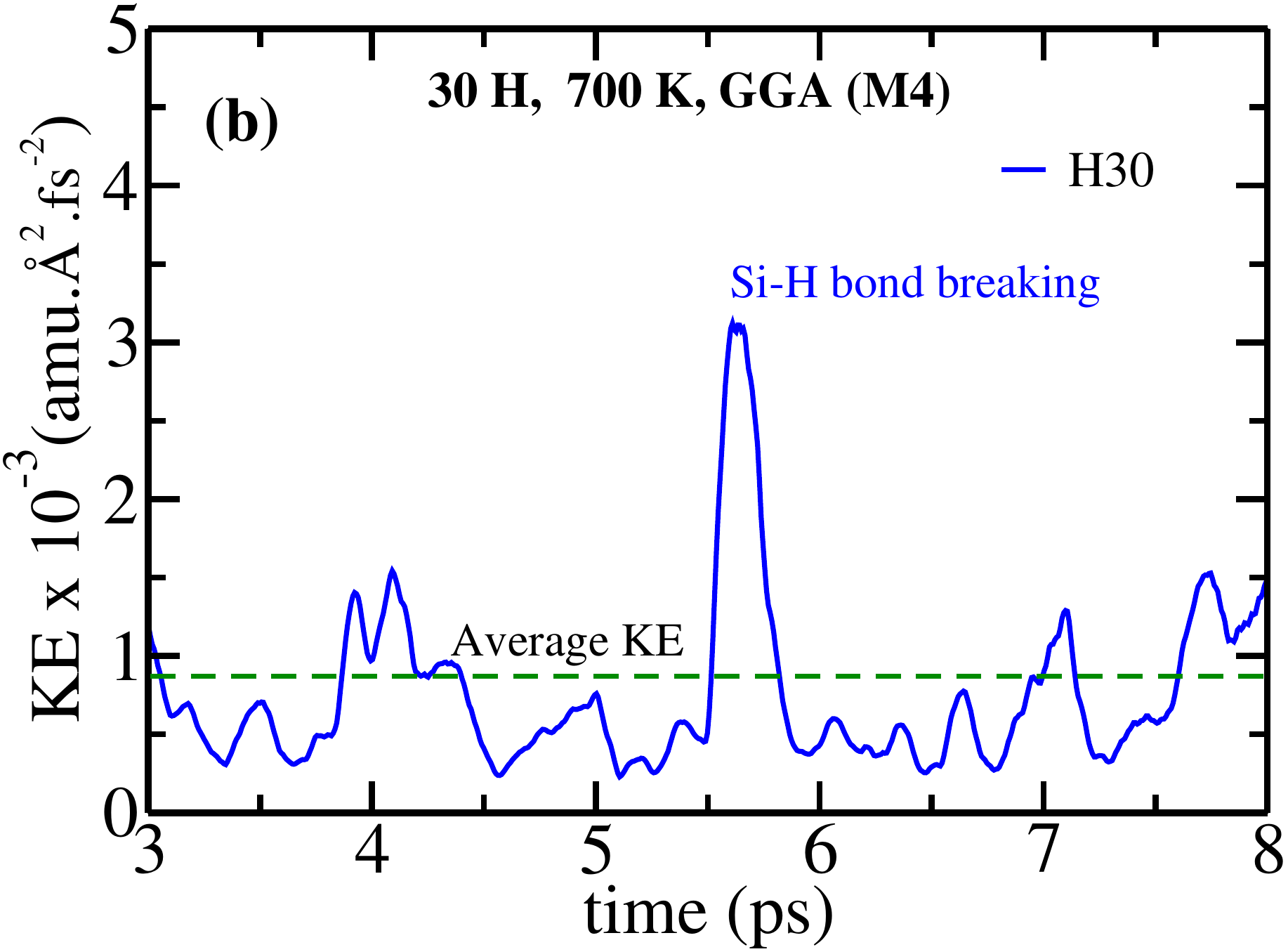}
\caption{
The evolution of the kinetic energy (KE) of two H 
atoms, H2 and H30, in monohydride Si--H bonds at 
700 K.  (a) The KE of H2 atom is found to stay 
below the average KE at 700 K for most of the 
simulation time. 
(b) The Si--H bond involving the high-energy H30 
atom dissociates near 5.5 ps and forms a new 
Si--H bond shortly thereafter.  The average KE 
of H atoms at 700 K is indicated as a dashed 
horizontal line (green).  The KE values correspond 
to a (central) moving average over a window 
of 100 fs. 
}
\label{KE}
\end{figure} 

\subsection{Kinetics of Si--H bond formation and 
dissociation on the surface of voids in {\asi}}

We now address a question of considerable importance 
concerning the kinetics of Si--H bond formation and 
dissociation on the surface of voids in {\asi}. 
While a complete understanding of these events requires 
the computation of bond formation and dissociation 
energies, and the knowledge of the potential-energy 
surface associated with Si--H bonding configurations, 
the problem can be addressed approximately by 
considering the kinetic energy of H atoms in 
AIMD simulations. 

Earlier, in Fig.~\ref{FIG3}, we have seen 
that the mean-square displacement (MSD) of H atoms inside 
the voids fluctuates within the range of 17--22 
{\AA}$^2$, depending upon the LDA or GGA. The MSD 
values are indicative of the fact that the great majority 
of H atoms stay bonded to Si atoms on the void surface 
as the time evolution of the system continues. However, 
it has been observed that a few H atoms in Si--H bonds 
can break free of the surface and move through the 
cavity to form new Si--H bonds 
at nearby active Si sites. This behavior is particularly pronounced 
at 700 K for a hydrogen load of 30 H atoms/void, where 
the presence of few H atoms with high kinetic energy 
makes it possible to dissociate an existing Si--H bond 
and form a new Si--H bond in the vicinity of the void 
surface. This is 
illustrated in Fig.~\ref{FIG7}, which shows the snapshots 
of two silicon-hydrogen bonding configurations in the 
vicinity of the void in M4, obtained from the GGA at 
3 ps and 18 ps. Figure \ref{FIG7}(a) 
shows a set of four Si--H bonds, involving H12, H16, H24, 
and H30, indicated as green-red pairs, at 3 ps, which 
are found to dissociate later during simulation.  
The breaking of Si--H bonds is reflected in Fig.~\ref{FIG7}(b), 
which shows that the H atoms (red) resulted from the 
dissociation of four Si--H bonds moved through the void 
and formed new Si--H bonds.  The latter are shown as 
yellow-red pairs, with the exception of one where 
a dissociated H atom (red) is found to be bonded with 
an active Si site (green). 
\begin{figure}[t!]
\centering
\includegraphics[width=0.35\textwidth]{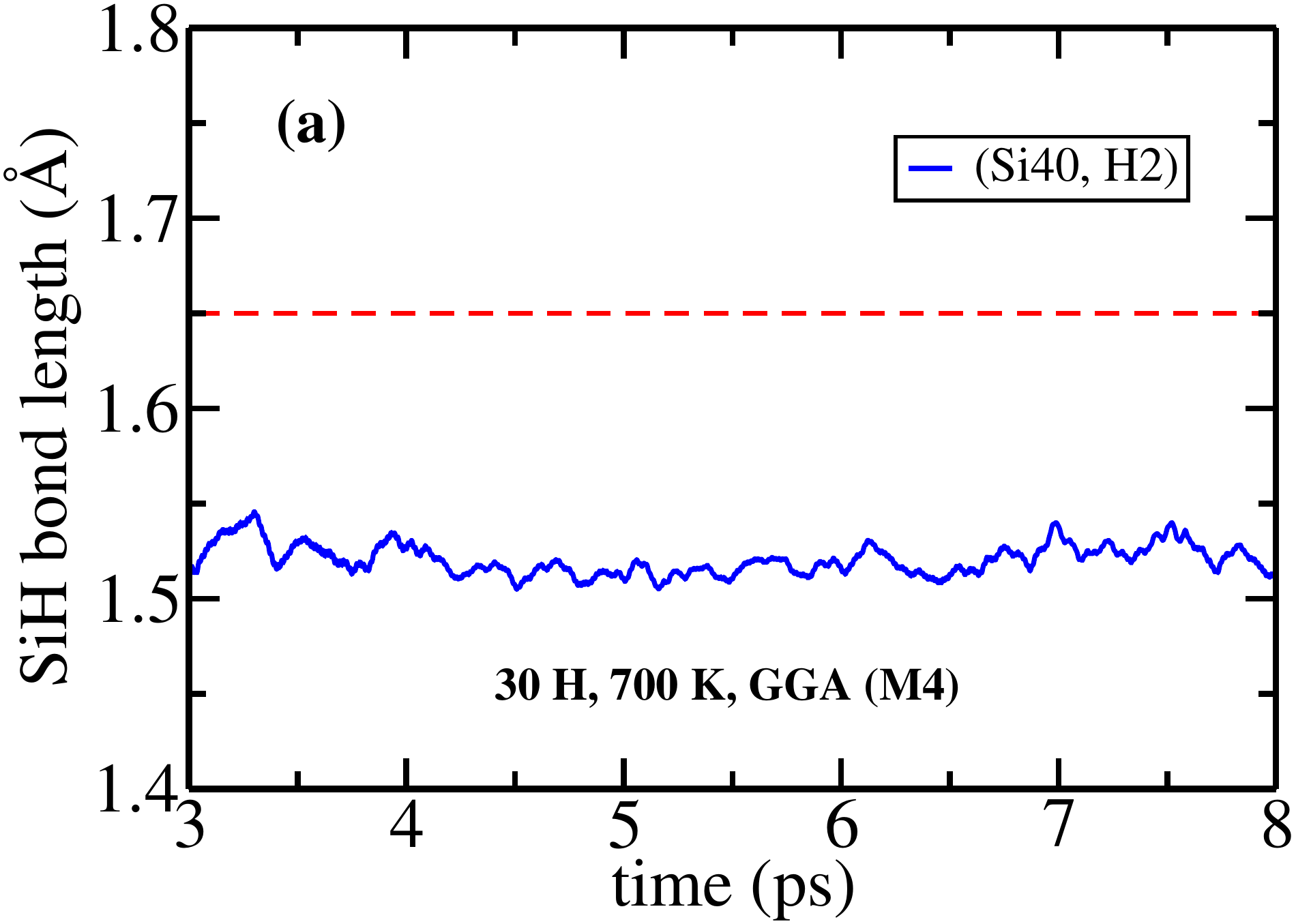}
\includegraphics[width=0.35\textwidth]{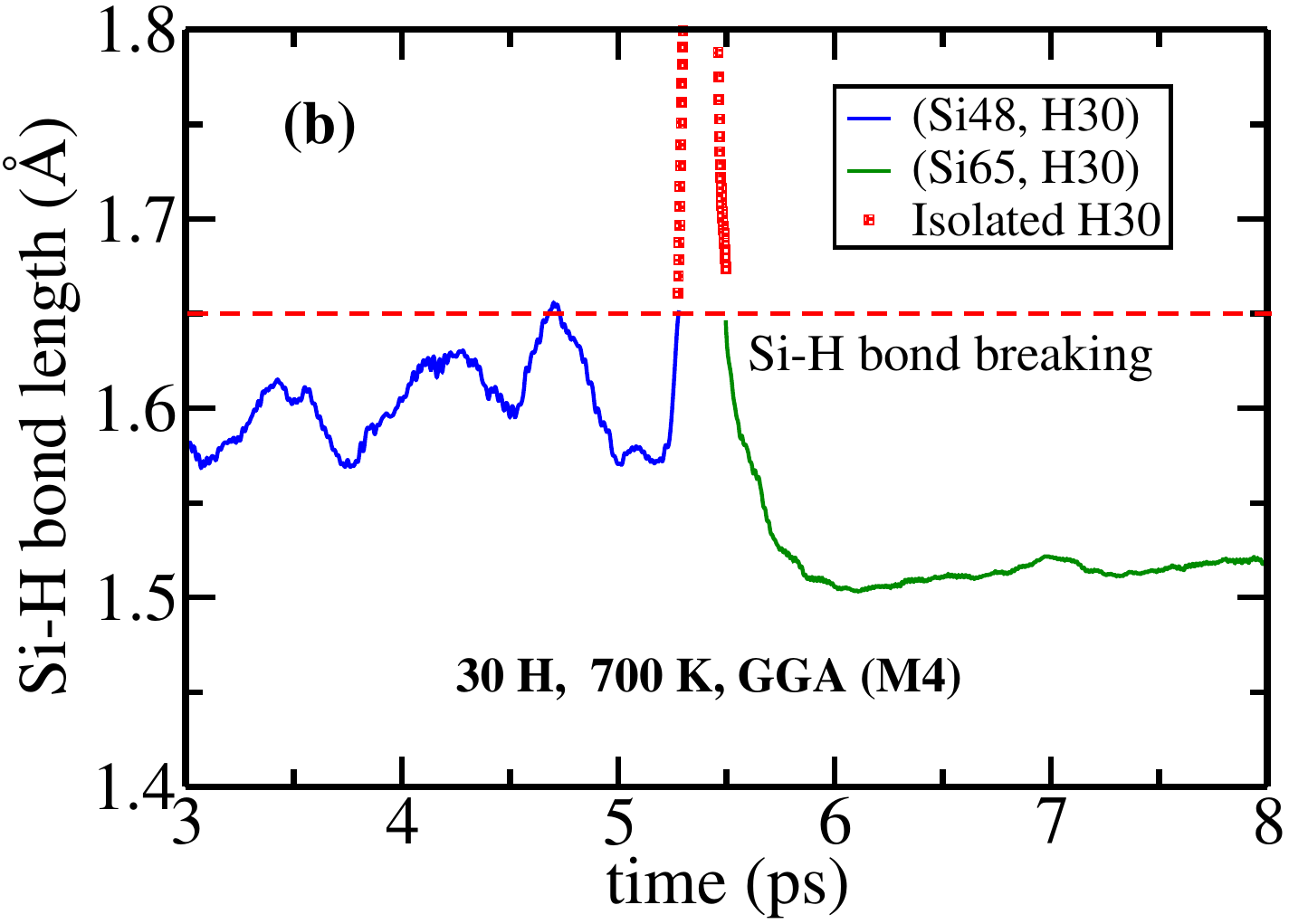}
\caption{ 
(a) The variation of the bond length of a stable
monohydride pair (Si40, H2) with time. (b) The 
dissociation of a monohydride Si--H bond during 
MD simulations. The Si--H bond associated with 
the pair (Si48, H30) breaks at 5.35 ps, which is 
followed by the formation of a new pair (Si65, H30) 
at 5.56 ps. The isolated state of H30 atom is 
indicated by red squares, where the `bond length' 
refers to the distance between H30 and the nearest 
Si atom. The cutoff value of the Si--H bond length, 
1.65 {\AA}, is indicated by a dashed horizontal 
line (red). 
}
\label{break}
\end{figure} 
The dissociation of Si--H bonds at 700 K can be 
understood from a kinetic point of view.  
Figure \ref{KE} shows the time evolution of the 
kinetic energy (KE) of two H atoms, H2 and H30, 
bonded to silicon atoms, Si40 and Si48, 
respectively, as monohydride Si--H bonds.  As 
the simulation proceeds, the KE values fluctuate 
around the average translational KE value of 
3$k_B$T/2 at temperature $T$, where $k_B$ is 
the Boltzmann constant.  However, the large 
kinetic energy of H30 atom at around 5.5 ps, which 
is about three times larger than the average 
KE at 700 K, leads to a rupture of the Si--H 
bond and subsequent formation of a new Si--H 
bond with a silicon atom in the vicinity of 
the void. By contrast, H2 forms a stable 
Si--H bond due its low kinetic energy 
during time evolution of the system. We have verified that 
the remaining three H atoms -- H12, H16, and 
H24 -- exhibit a similar behavior as H30 as 
far as the KE values are concerned. 

The dissociation of Si--H bonds due to high KE 
values of H atoms can be observed by tracking 
the Si--H bond length of a dissociated pair with 
time.  Figure \ref{break} shows the evolution 
of Si--H bond lengths associated with the pairs 
(Si40, H2) and (Si48, H30). The pair (Si48, H30) 
breaks at 5.35 ps, which is followed by the formation of a new pair 
(Si65, H30) at 5.56 ps, but the low-energy pair 
(Si40, H2) continues to stay bonded during its evolution. 
The variation of the bond 
length of the (dissociated) pair before and 
after dissociation are shown in 
Fig.~\ref{break}(b). The transient isolated state of H30, between 
5.35 ps and 5.56 ps, is shown as red 
squares, where the `bond length' corresponds to the 
distance between H30 and the nearest Si atom 
in the void region. It goes without saying that 
the dissociation and subsequent formation of 
Si--H bonds on the void surface do not affect 
the MSD values of the H atoms as long as the 
H atoms stay within the cavity through the 
formation of new bonding configurations. 

\begin{table}[t!]
\caption{\label{TAB1}
Statistics of bonded and non-bonded hydrogens inside voids 
for varying hydrogen loads.  The tabulated values indicate 
the amount of H atoms (in per cent of total H) that resides 
in voids as isolated hydrogen (H$_{\text{iso}}$), H$_2$ molecules, 
and SiH/SiH$_2$ bonds. Hydrogen atoms outside the void 
region are listed as Ex$_H$. Asterisks denote the actual 
number of isolated H atoms. 
}
\vskip 0.2cm 
\centering 
\begin{ruledtabular} 
\begin{tabular}{c|c|c|r|r|r|r}
H load & XC & H$_{\text{iso}}$ &H$_{2}$  & SiH& SiH$_{2}$ &Ex$_H$ \\
\hline
\multicolumn{7}{c}{400 K} \\
\hline
10      & GGA   & 0  & 15.0 & 67.50 & 0  & 17.50\\
        & LDA   & 0  & 10.0 & 82.50 & 0  & 7.50\\

20      & GGA   & 0  & 32.50 & 47.50 & 10.0 & 10.0\\
        & LDA   & 0  & 25.0 & 62.75 & 5.0  & 7.50\\

30      & GGA   & 0  & 45.0 & 40.0 & 11.67 & 3.33\\
        & LDA   & 0  & 28.33 & 47.50 & 18.33 & 5.83\\
\hline
\multicolumn{7}{c}{700 K} \\
\hline
10      &GGA        & 0  & 25.0 & 55.0 & 0  & 20.0\\
        &LDA        & 0  & 0  & 77.50 & 5.0  & 17.50\\

20      &GGA        & 0  & 27.50 & 45.0 & 12.50 & 15.0\\
        &LDA        & 0  & 17.50 & 66.25 & 2.50  & 13.75\\

30      &GGA        & 0.83 (1$^{*}$)  & 35.56 & 41.11 & 8.89  & 13.33\\
        &LDA        & 1.65 (2$^{*}$)  & 16.67 & 56.67 & 11.67 & 13.33
\end{tabular} 
\end{ruledtabular} 
\end{table}

\subsection{Hydrogen microstructure and morphology of voids in {\asi}}
The discussion in the preceding sections so far is mostly 
confined to the dynamics of H atoms inside the voids and to what extent
the motion of H atoms is affected by the exchange-correlation 
approximation and the hydrogen load inside the voids, as far 
as the MSD of H atoms is concerned. We now examine the 
microstructure of hydrogen distributions on void surfaces 
and the morphological character of the voids, which result 
from the movement of hydrogen and silicon atoms in the vicinity 
of the voids for a varying hydrogen load. 
In particular, we discuss the formation of various 
bonded hydrogens (BH) and non-bonded hydrogens (NBH), as 
well as the restructuring of the void surfaces 
during annealing at 400 K and 700 K, with an increasing 
concentration of H atoms within voids, and the dependence 
of hydrogen microstructure on the LDA and GGA. The BH 
and NBH play an important role in characterizing the 
structural and optical properties of 
{\asih}~\cite{Biswas2011,Sekimoto2016,Sekimoto2018}. 
These properties can be studied experimentally using an array 
of experiments, such as positron-annihilation lifetime 
(PAL) spectroscopy~\cite{Sekimoto2016}, Rutherford backscattering 
spectrometry (RBS)~\cite{Sekimoto2018}, hydrogen-effusion 
measurements~\cite{Beyer2003,Beyer1982}, and 
Fourier-transform infrared spectroscopy-attenuated total 
reflections (FTIR-ATR)~\cite{Sekimoto2014}.

\begin{figure}[t!]
\begin{center} 
\includegraphics[width=0.325\textwidth]{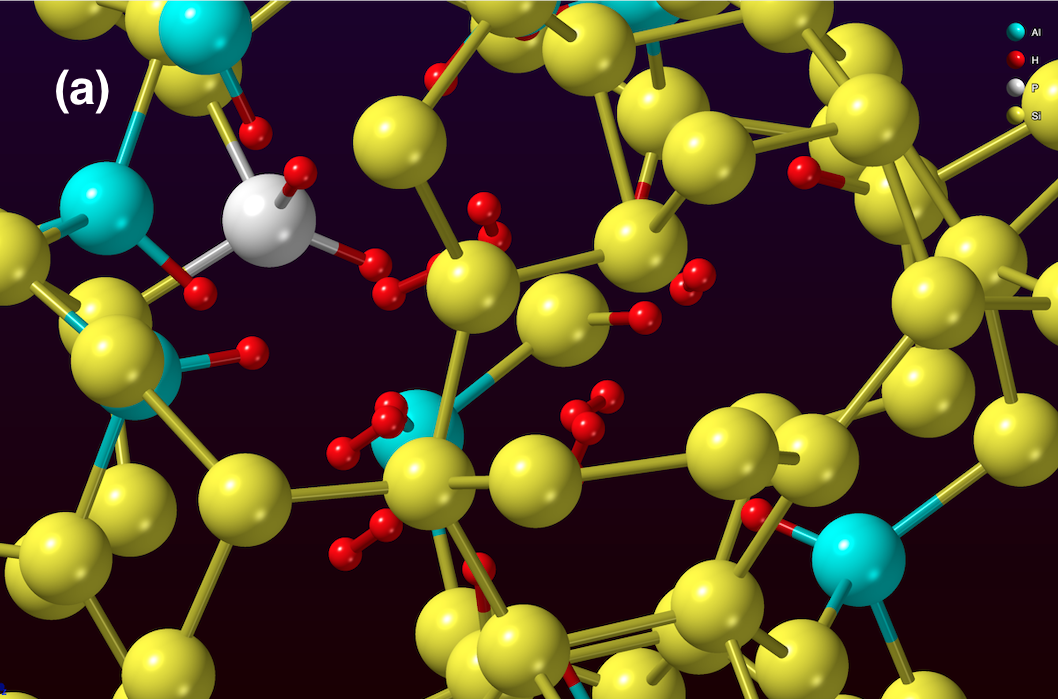}
\includegraphics[width=0.325\textwidth]{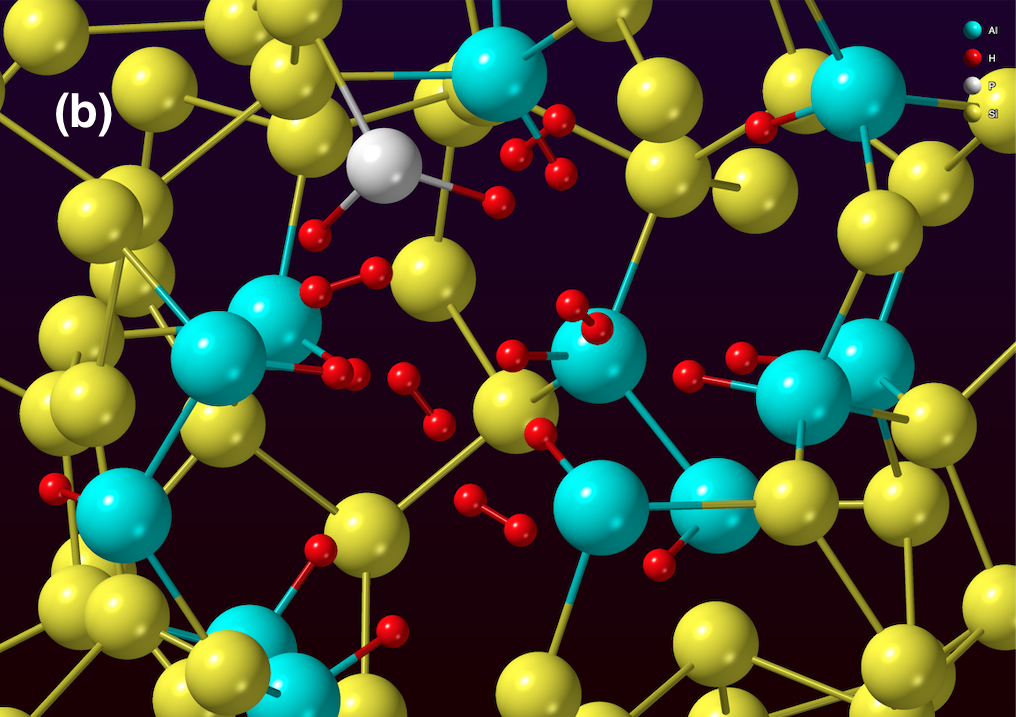}
\caption{
\label{FIG10}
Hydrogen-bonding configurations, including H$_2$ molecules, 
in a void-surface region of radius 8 {\AA} for a hydrogen 
load of 30 H atoms/voids at (a) 400 K and (b) 700 K from 
the GGA in model M4. Silicon atoms associated with monohydride 
(SiH) and dihydride (SiH$_2$) configurations are shown 
in cyan and white colors, respectively, whereas bonded H 
atoms and molecules are shown in red color. 
}
\end{center} 
\end{figure}

\begin{figure}[t!]
\centering
\includegraphics[width=0.35\textwidth]{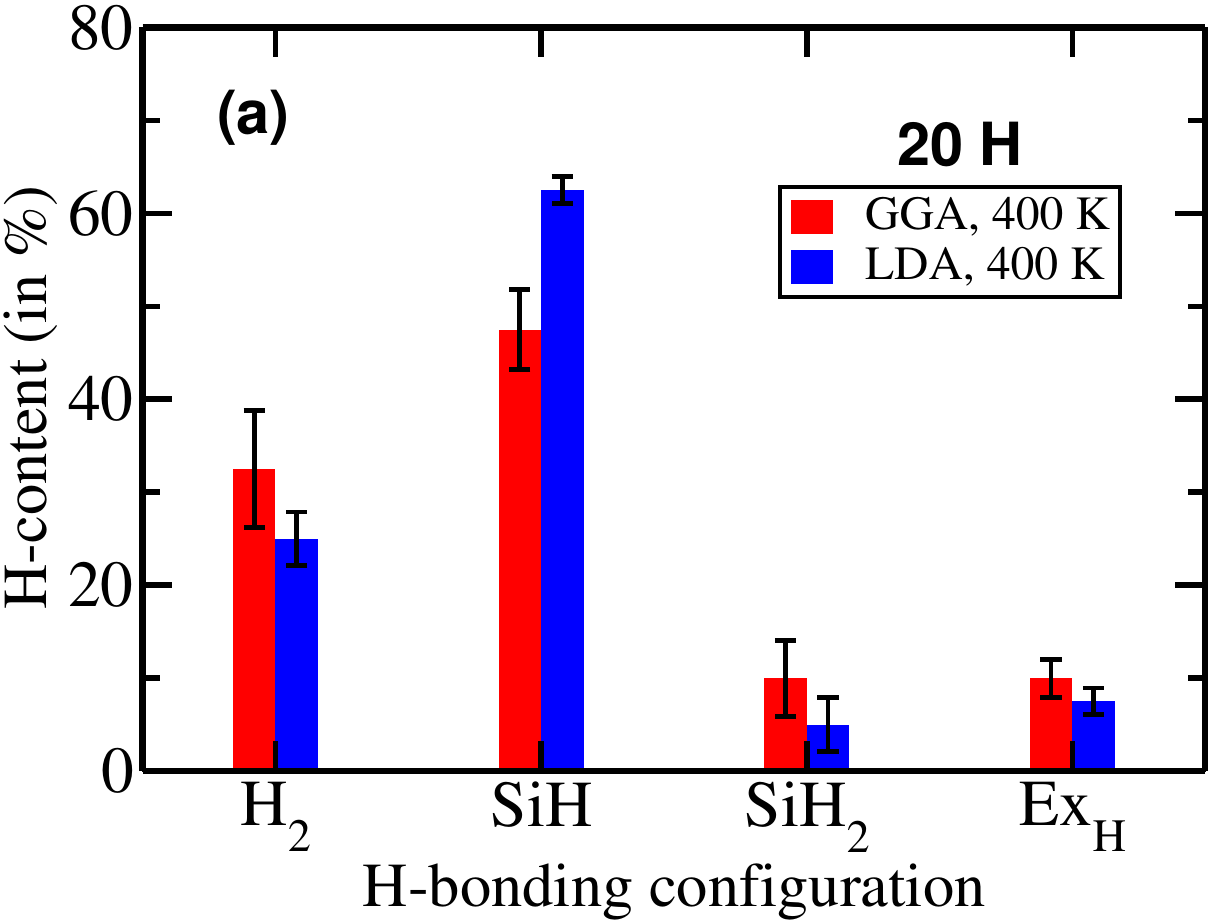}
\includegraphics[width=0.35\textwidth]{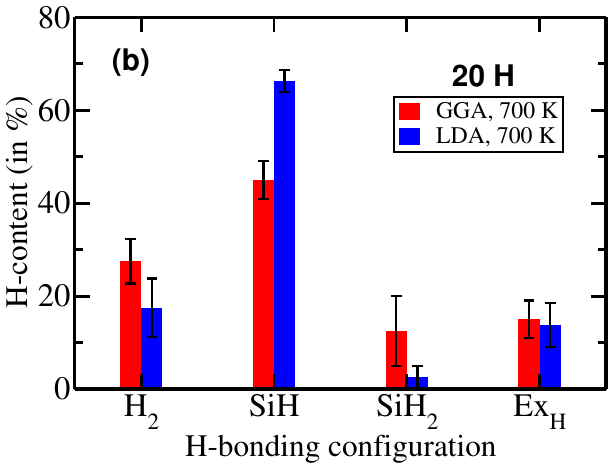}
\caption{\label{FIG11}
Histograms showing the content of H atoms (in 
at.\,\%) in bonded and non-bonded hydrogens 
in the vicinity of voids in {\asi} for a 
hydrogen load of 20 H atoms/voids at 400 K and 700 K. 
The results are obtained by averaging over 
all models. 
}
\end{figure}

Table \ref{TAB1} lists the statistics of various silicon-hydrogen 
bonding configurations and non-bonded hydrogens, e.g., H$_2$ 
molecules and one or two isolated H atoms, near the void, defined 
by a spherical region of radius 8 {\AA}. 
The presence of isolated H atoms in the network is an 
artifact (of simulations), which arises from our 
choice of the cutoff value of 1.65 {\AA} for Si--H bonds. It 
has been observed that the isolated H atoms at 700 K, 
listed in Table \ref{TAB1},  are at a distance of 
1.66 {\AA} from the nearest Si atom (for the GGA) 
and at distances of 1.67 {\AA} and 1.7 {\AA} from the 
neighboring Si atoms (for the LDA). 
The results from Table \ref{TAB1} and an analysis of 
the void surfaces for the hydrogen load of 10, 20, 
and 30 H atoms/void show that the surface of the 
voids is mostly decorated with monohydride Si--H bonds. 
A few dihydride Si--H$_2$ bonds are also spotted for 
a hydrogen load of 20 and 30 H atoms/void, and a 
notable amount of hydrogen can be seen to appear 
inside the voids as H$_2$ molecules, especially for 
hydrogen loads of 20 and 30 H atoms/voids. It is also 
evident from Table \ref{TAB1} that, in comparison 
with the GGA, the LDA overestimates the number of 
SiH bonds but underestimates the count of H$_2$ 
molecules. The presence of few SiH$_2$ bonding 
configurations makes it difficult to comment on 
the dependence of SiH$_2$ configurations on the 
XC approximation from the available data. 
Figure \ref{FIG10} shows a three-dimensional 
rendering of Si--H and Si--H$_2$ bonds in the 
vicinity of the void surface in M4, along with 
a few H$_2$ molecules for a hydrogen load of 
30 H atoms/voids at 400 K and 700 K.  

Experimental studies using infrared measurements by Chabal 
and Patel~\cite{Chabal1984} suggest that the number 
density of H$_2$ molecules in nanometer-size voids in {\asi} 
is of the order of 10$^{21}$ cm$^{-3}$. This observation is 
found to be consistent with the values listed in Table 
\ref{TAB1}, which can be roughly translated into 3--13 $\times$ 
10$^{21}$ cm$^{-3}$ for the GGA and 2--8 $\times$ 
10$^{21}$ cm$^{-3}$ for the LDA at 400 K, assuming a 
spherical void of radius 5--8 {\AA} and a hydrogen load 
of 30 H atoms/void.  None of the AIMD runs in this study 
showed any isolated H atoms within the voids, except one 
or two at 700 K, which are statistically insignificant. 
As discussed earlier, a few highly mobile H atoms are 
found to diffuse out of the void region due to high 
KE values of these atoms. The number 
of such atoms are listed as Ex$_H$ in Table \ref{TAB1}. 
Figure \ref{FIG11} summarizes the results from Table 
\ref{TAB1}, by showing the number of Si--H, Si--H$_2$, 
and H$_2$ molecules for a hydrogen load of 20 H 
atoms/void at 400 K and 700 K.

\begin{figure}[t!]
\centering
\includegraphics[width=0.4\textwidth]{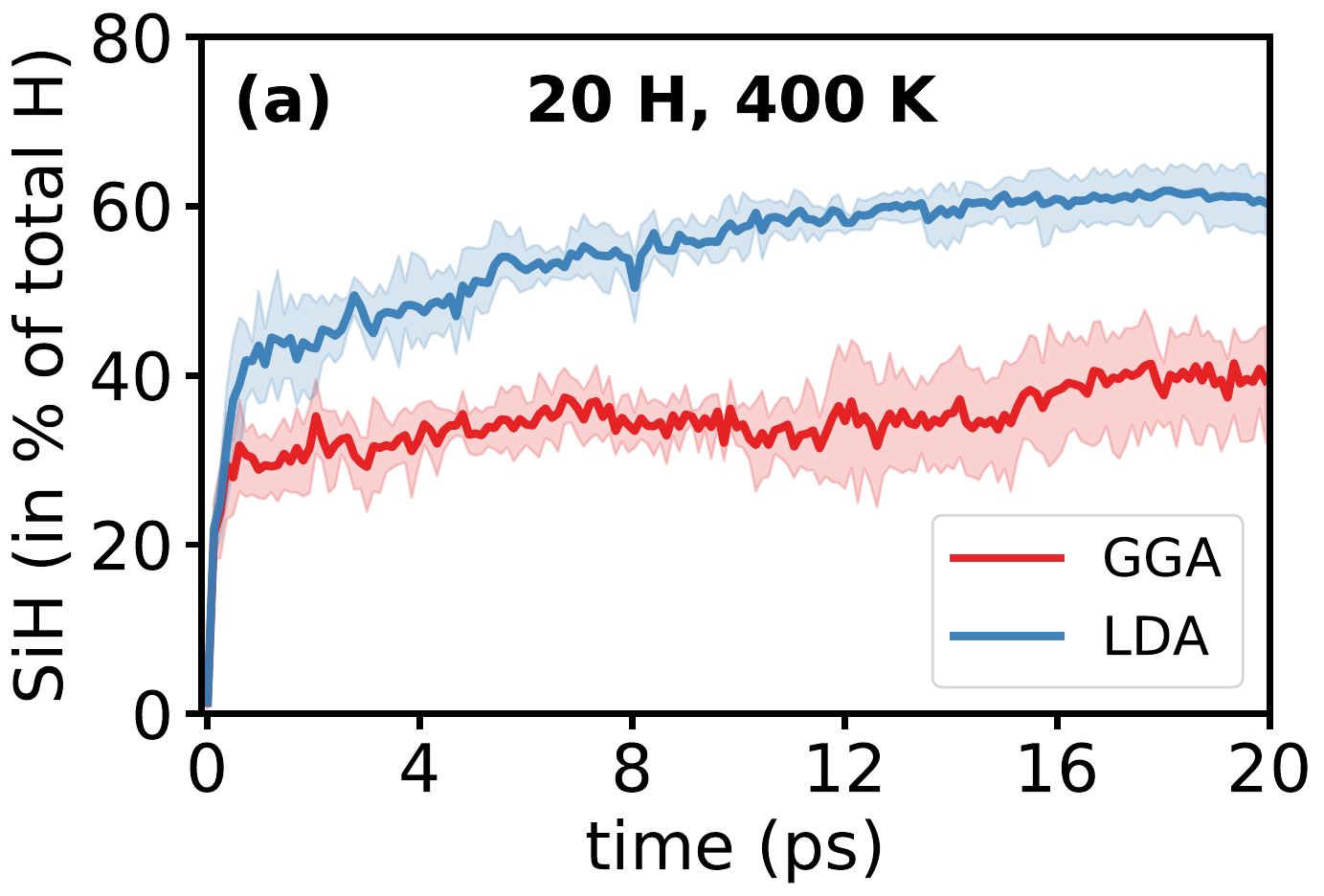}
\includegraphics[width=0.4\textwidth]{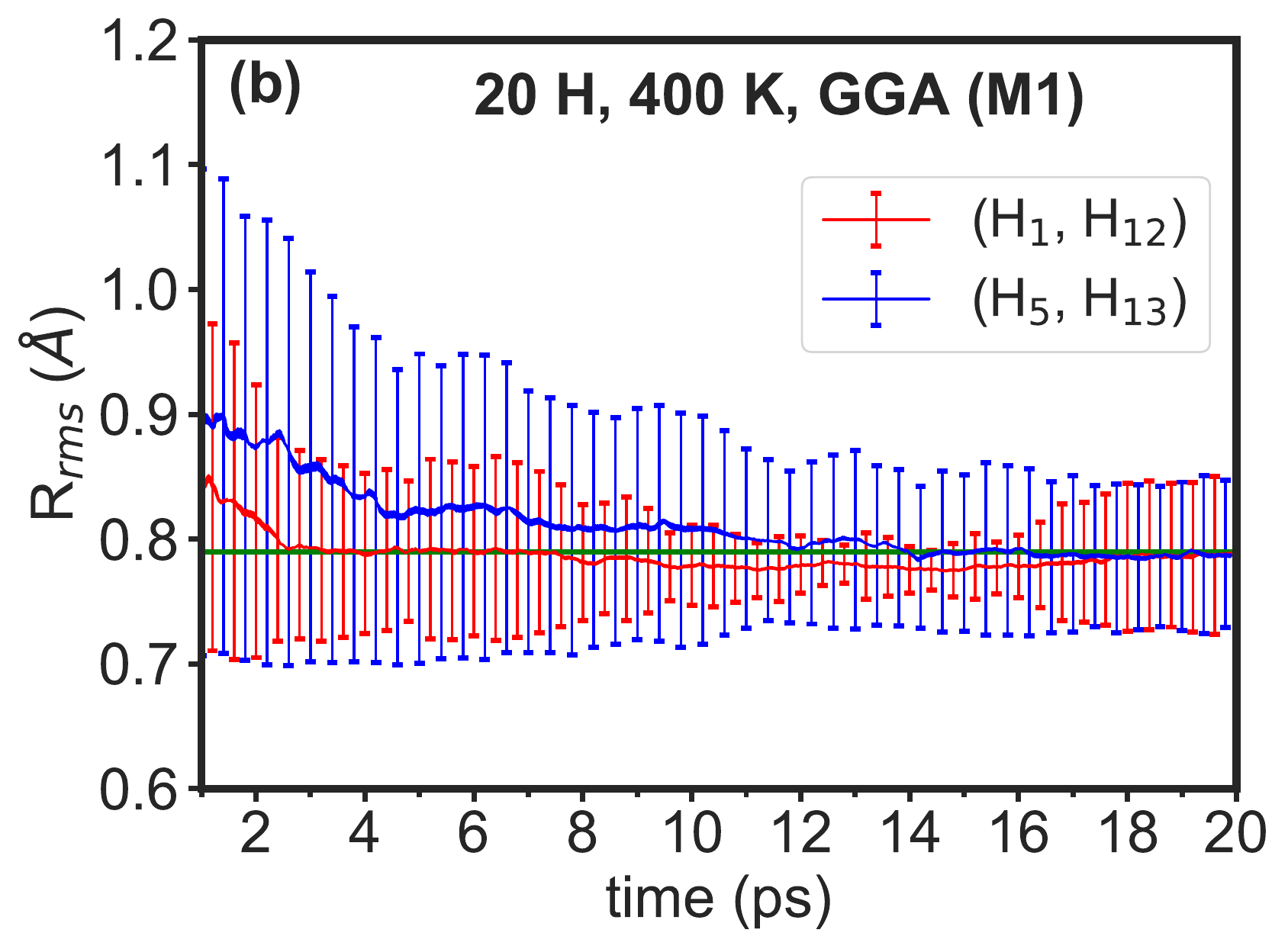}
\caption{\label{FIG12}
(a) The evolution of monohydride Si--H bonds 
at 400 K in the GGA and the LDA for a hydrogen load 
of 20 H atoms/void. The $y$ axis indicates the amount 
of hydrogen atoms (in per cent of total H atoms) 
that resides in Si--H bonds. (b) The formation of H$_2$ 
molecules within a void at 400 K in the GGA for 
a hydrogen load of 20 H atoms/voids. The 
RMS distance and its standard deviation 
between two pairs of H atoms, averaged over a moving 
window of width 200 fs, are shown against 
time.  
}
\end{figure}

Since the restructuring of void surfaces is largely 
characterized by Si--H bonds on the walls 
of the voids, it is instructive to examine the formation 
of these bonds during the course of simulation. Figure 
\ref{FIG12}(a) shows the time evolution of the hydrogen 
content of SiH bonds (in per cent of total H) for a 
hydrogen load of 20 H atoms/voids in the GGA and the 
LDA at 400 K. The formation of Si--H bonds begins very 
rapidly within the first few picoseconds and then 
it gradually converges by 18 ps.  The LDA seems to 
overestimate the number of SiH bonds by about 32\%, 
compared with the same from the GGA (cf.\,Table \ref{TAB1}). 
This notable difference between the LDA and the 
GGA results indicates the need for choosing an accurate 
XC functional and a long simulation time for studying 
the microstructure of 
SiH and SiH$_2$ on the walls of the voids. Likewise, 
the formation of H--H pairs, or H$_2$ molecules, 
during annealing is illustrated in Fig.~\ref{FIG12}(b), 
by plotting the evolution of the RMS distance 
between two pairs of H atoms at 400 K in the GGA for 
a hydrogen load of 20 H atoms/voids in M1. The first H$_2$ 
molecule, consists of (H1, H12), was formed within 
the first 3 ps, whereas the second one, (H5, H13), 
was formed at around 14 ps.  The RMS distance between 
the pair of H atoms and the corresponding standard 
deviation are obtained by averaging over a moving time 
window of width 200 fs.  

\begin{figure}[t!]
\centering
\includegraphics[width=0.235\textwidth]{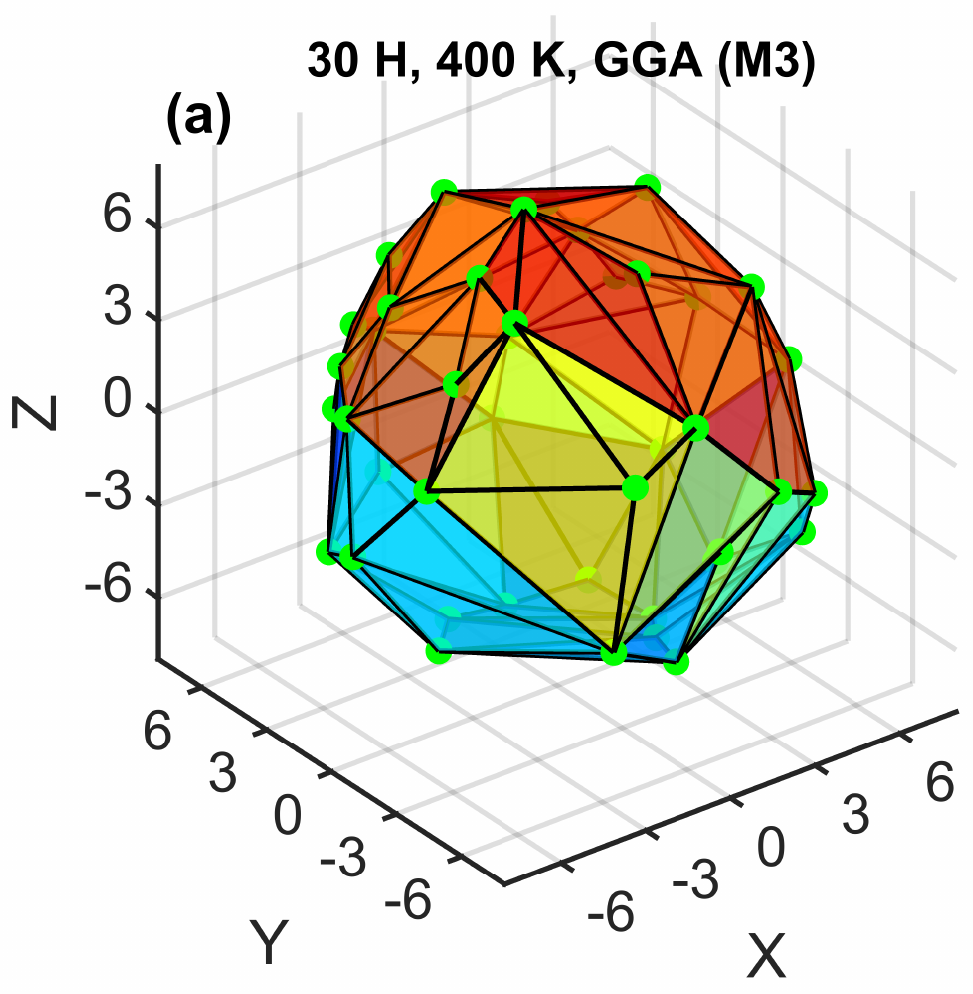}
\includegraphics[width=0.235\textwidth]{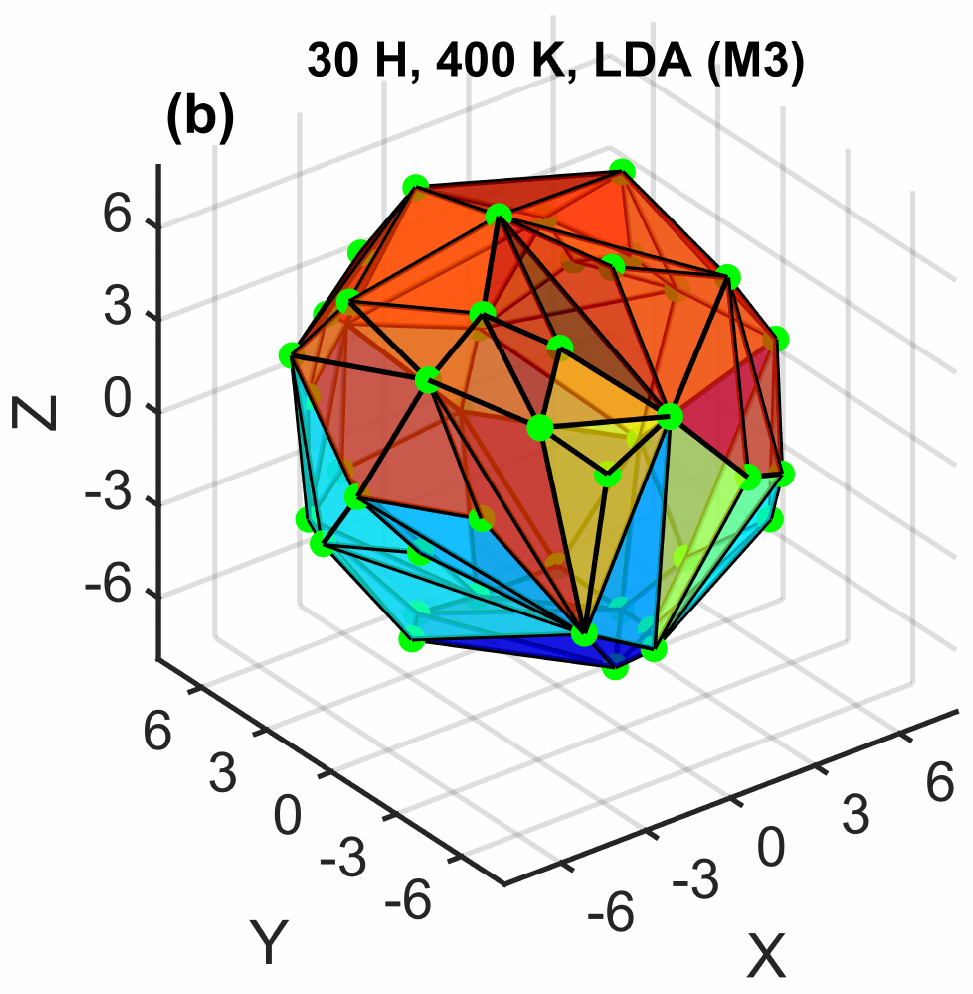}
\includegraphics[width=3.5cm, height=3.55cm]{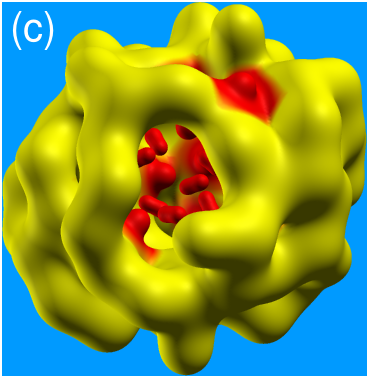}
\includegraphics[width=3.5cm, height=3.55cm]{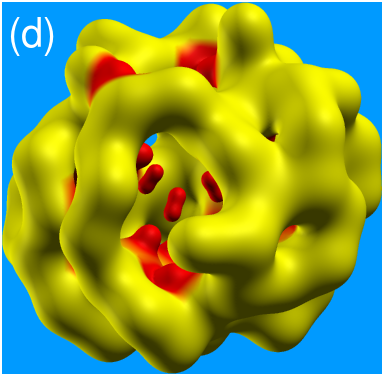}
\caption{\label{FIG13} 
The reconstruction of a three-dimensional shape of 
a void (in M3) from two sets of void-surface atoms 
at 400 K for a hydrogen load of 30 H atoms/void. 
The convex polyhedra obtained from the void-surface 
atoms in (a) the GGA and (b) the LDA, using the 
convex-hull approximation. The corresponding non-convex 
surfaces using the same set of void-surface atoms 
are shown in (c) and (d), respectively. 
Silicon-hydrogen bonds and H$_2$ molecules are shown 
in yellow-red and red-red colors, respectively. 
}
\end{figure}

We conclude this section by making the following observation 
on the linear size and the shape of the voids obtained 
from annealing and total-energy relaxations in the presence 
of H atoms. While the linear size of a void 
can be estimated from the radius of gyration of a set 
of atoms, which define the void surface, the reconstruction 
of a three-dimensional shape of a void 
from a finite set of atomic positions is a nontrivial 
problem.  A somewhat crude but simple and useful approach 
is to approximate the void shape by constructing the 
minimal convex polyhedron, or a convex hull, formed by 
the set of void-surface atoms. The approach assumes that any 
restructuring of the void surface -- caused 
by the movement of the void-surface atoms -- would 
be reflected in the shape of the convex hull, which 
is associated with the void region. However, 
it has been observed~\cite{BiswasIOP2015} that 
the actual shape of voids can be rather complex 
and non-convex in nature, and it may not be 
represented accurately by a convex polyhedron. An 
approximate non-convex surface can be constructed 
via convolution of the position of the void-surface 
atoms using three-dimensional Gaussian basis 
functions and choosing a suitable value of the 
isosurface parameter for the atomic pseudo-surface 
of Si and H atoms. The {\sc Xcrysden}~\cite{xc} 
package can generate such pseudo-surfaces. 
A more general discussion on the reconstruction 
of such non-convex void shapes can be found in 
Refs.~\cite{Biswas2020,BiswasIOP2015}. 

\begin{table}[t!]
\caption{\label{TAB2}
Linear (\AA) and volumetric (\AA$^3$) measures of the voids, reconstructed from 
the convex-hull approximation. $R_G$, $R_H$, and $V_H$ indicate 
the radius of gyration, the convex-hull radius, and the volume 
of the hull, respectively. $S$ and $N_H$ represent the sphericity 
and the number of atoms on the convex hull, respectively.
}
\vskip 0.2cm 
\centering 
\begin{ruledtabular}
\begin{tabular}{c|c|c|c|c|c|c}
H load & XC & R$_G$ & R$_H$ &V$_H$   & $S$ & $N_H$ \\
\hline
\multicolumn{7}{c}{400 K} \\
\hline
10      & GGA   & 6.392  & 7.162        & 1275.04       & 0.884 & 41    \\
        & LDA    & 6.517 & 7.205    & 1305.45       & 0.888 & 41    \\
%        & Harris & 6.540 & 7.234        & 1322.68       & 0.883 & 41    \\

20      & GGA   & 6.416  & 7.204        & 1295.95       & 0.883 & 41\\
        & LDA    & 6.566 & 7.218    & 1316.80       & 0.890 & 42    \\
%        & Harris & 6.552 & 7.230        & 1320.30       & 0.885 & 43    \\

30      & GGA   & 6.489 & 7.231 & 1318.42       & 0.885 & 42    \\
        & LDA    & 6.541 & 7.263    & 1332.45       & 0.887 & 42    \\
%        & Harris & 6.592 & 7.292        & 1350.73       & 0.887 & 42            \\
\hline
\multicolumn{7}{c}{700 K} \\
\hline
10  & GGA  & 6.192 & 7.102      & 1242.36       & 0.883 & 40            \\
    & LDA    & 6.347 & 7.175    & 1287.78       & 0.887 & 41    \\
%    & Harris & 6.395 & 7.247        & 1305.67       & 0.880 & 38    \\

20  & GGA        & 6.365 & 7.242 & 1299.63       & 0.880 & 39    \\
    & LDA    & 6.461 & 7.274    & 1323.15       & 0.879 & 39    \\
%    & Harris & 6.411 & 7.238        & 1320.64       & 0.890 & 39    \\

30  & GGA         & 6.430 & 7.202        & 1319.43       & 0.892 & 42 \\
    & LDA    & 6.475 & 7.238    & 1317.39       & 0.882 & 41    \\
%    & Harris & 6.435 & 7.278        & 1333.39       & 0.881 & 40\\
\end{tabular}
\end{ruledtabular}
\end{table} 

Figure \ref{FIG13} shows the approximate shape 
of a void with a hydrogen load of 30 H atoms/void 
(in model M3) obtained from annealing at 400 K, 
followed by total-energy optimization. The convex 
polyhedra shown in Figs.~\ref{FIG13}(a) and 
\ref{FIG13}(b) correspond to the set of void-surface atoms 
obtained from the GGA and the LDA, respectively. 
The respective non-convex shapes of the void, 
obtained via the convolution of the same set of 
void-surface atoms using the Gaussian functions 
centered at the atomic sites, are shown in 
Figs.~\ref{FIG13}(c) and \ref{FIG13}(d).  Although 
the polyhedra in Figs.~\ref{FIG13}(a)-(b) appear 
different, the difference is not particularly 
noteworthy as far as the hull radius, the 
convex-hull volume, and the sphericity of the 
polyhedra are concerned. These values are listed 
in Table \ref{TAB2}.  The sphericity, $S$, 
of an object is defined as the ratio of the 
surface area of a sphere, $A_s$, to that of 
the object, $A$, both having an identical 
volume $V$. This definition leads 
to~\cite{Wadell1935},  
\[
S = \frac{A_s}{A} = \frac{\pi^{\frac{1}{3}}(6V)^{\frac{2}{3}}}{A}.
\]
Here, we have used the volume and the corresponding 
surface area of the convex hull of a void surface to 
estimate $V$ and $A$, respectively. A review of $S$ 
and other values in Table \ref{TAB2} 
suggests that the LDA marginally overestimates 
the volume of the void, which is evident from 
the gyrational and hull radii, and the hull 
volume of the voids for all hydrogen loads and 
temperatures.  This observation is also consistent 
with the somewhat smaller values of the MSD of 
H atoms (within voids) that we have observed 
for the GGA calculations in Figs.~\ref{FIG3}(a)-(b). 

\vspace*{0.5cm} 
\section{Conclusions}
In this paper, we have studied {\it ab initio} 
dynamics of hydrogen atoms inside voids in {\asi} 
with an emphasis on hydrogen diffusion and 
the resulting structure of the void surfaces 
with respect to a varying concentration of 
hydrogen at 400 K and 700 K. A comparison of the results 
obtained from the LDA and the GGA reveals that 
the former considerably overestimates the number 
of monohydride Si--H bonds but underestimates 
the presence of H$_2$ molecules inside the 
cavities, irrespective of the annealing 
temperature and the concentration of hydrogen. 
The surfaces of the voids are found to be 
primarily passivated with monohydride Si--H bonds 
and a few dihydride SiH$_2$ bonds at high 
concentration of hydrogen. Neither the LDA nor the GGA shows any 
presence of SiH$_3$ configurations even for 
a high concentration/load of 30 H atoms/void. 
The number densities of the bonded and non-bonded 
hydrogens observed in this study are found to be 
consistent with those from the infrared and 
Rutherford back scattering (RBS) measurements. 
The study reveals that the kinetics of Si--H bond 
formation and dissociation during AIMD simulations 
can be approximately described and understood 
by considering the (translational) kinetic energy 
of H atoms inside the voids. Hydrogen atoms with 
KE values significantly higher than the 
average KE of the system at a given temperature 
are found to dissociate from Si--H bonds on 
the surface of the voids. The resulting isolated H 
atoms then diffuse through the void region to 
form new bonds with nearby active Si atoms within 
a fraction of a picosecond in our simulations.  
The results also show that a somewhat higher value of 
the mean-square displacement of the H atoms 
within voids in the LDA can be attributed to 
the reconstruction of the void surface through 
the formation of Si--H bonds. This is also 
reflected in the linear size of the voids obtained 
from the convex-hull approximation.  

\section*{acknowledgments}
The work was partially supported by the U.S. 
National Science Foundation (NSF) under Grant 
No.\,DMR 1833035.  The authors acknowledge the 
use of computing resources at the University of 
Southern Mississippi, supported by the NSF 
under the Major Research Instrumentation (MRI) 
program via Grant No.~ACI 1626217. 

%\bibliographystyle{apsrev4-2}
%\bibliography{ref.bib}

\begin{thebibliography}{36}%
\makeatletter
\providecommand \@ifxundefined [1]{%
 \@ifx{#1\undefined}
}%
\providecommand \@ifnum [1]{%
 \ifnum #1\expandafter \@firstoftwo
 \else \expandafter \@secondoftwo
 \fi
}%
\providecommand \@ifx [1]{%
 \ifx #1\expandafter \@firstoftwo
 \else \expandafter \@secondoftwo
 \fi
}%
\providecommand \natexlab [1]{#1}%
\providecommand \enquote  [1]{``#1''}%
\providecommand \bibnamefont  [1]{#1}%
\providecommand \bibfnamefont [1]{#1}%
\providecommand \citenamefont [1]{#1}%
\providecommand \href@noop [0]{\@secondoftwo}%
\providecommand \href [0]{\begingroup \@sanitize@url \@href}%
\providecommand \@href[1]{\@@startlink{#1}\@@href}%
\providecommand \@@href[1]{\endgroup#1\@@endlink}%
\providecommand \@sanitize@url [0]{\catcode `\\12\catcode `\$12\catcode
  `\&12\catcode `\#12\catcode `\^12\catcode `\_12\catcode `\%12\relax}%
\providecommand \@@startlink[1]{}%
\providecommand \@@endlink[0]{}%
\providecommand \url  [0]{\begingroup\@sanitize@url \@url }%
\providecommand \@url [1]{\endgroup\@href {#1}{\urlprefix }}%
\providecommand \urlprefix  [0]{URL }%
\providecommand \Eprint [0]{\href }%
\providecommand \doibase [0]{https://doi.org/}%
\providecommand \selectlanguage [0]{\@gobble}%
\providecommand \bibinfo  [0]{\@secondoftwo}%
\providecommand \bibfield  [0]{\@secondoftwo}%
\providecommand \translation [1]{[#1]}%
\providecommand \BibitemOpen [0]{}%
\providecommand \bibitemStop [0]{}%
\providecommand \bibitemNoStop [0]{.\EOS\space}%
\providecommand \EOS [0]{\spacefactor3000\relax}%
\providecommand \BibitemShut  [1]{\csname bibitem#1\endcsname}%
\let\auto@bib@innerbib\@empty
%</preamble>
\bibitem [{\citenamefont {Street}(2009)}]{Street2009}%
  \BibitemOpen
  \bibfield  {author} {\bibinfo {author} {\bibfnamefont {R.~A.}\ \bibnamefont
  {Street}},\ }\href {https://doi.org/10.1002/adma.200803211} {\bibfield
  {journal} {\bibinfo  {journal} {Adv. Materials}\ }\textbf {\bibinfo {volume}
  {21}},\ \bibinfo {pages} {2007} (\bibinfo {year} {2009})}\BibitemShut
  {NoStop}%
\bibitem [{\citenamefont {Wolf}\ \emph {et~al.}(2012)\citenamefont {Wolf},
  \citenamefont {Descoeudres}, \citenamefont {Holman},\ and\ \citenamefont
  {Ballif}}]{HJT2012}%
  \BibitemOpen
  \bibfield  {author} {\bibinfo {author} {\bibfnamefont {S.~D.}\ \bibnamefont
  {Wolf}}, \bibinfo {author} {\bibfnamefont {A.}~\bibnamefont {Descoeudres}},
  \bibinfo {author} {\bibfnamefont {Z.~C.}\ \bibnamefont {Holman}},\ and\
  \bibinfo {author} {\bibfnamefont {C.}~\bibnamefont {Ballif}},\ }\href
  {https://doi.org/https://doi.org/10.1515/green-2011-0018} {\bibfield
  {journal} {\bibinfo  {journal} {Green}\ }\textbf {\bibinfo {volume} {2}},\
  \bibinfo {pages} {7 } (\bibinfo {year} {2012})}\BibitemShut {NoStop}%
\bibitem [{\citenamefont {Ge}\ \emph {et~al.}(2012)\citenamefont {Ge},
  \citenamefont {Ling}, \citenamefont {Wong}, \citenamefont {Mueller},\ and\
  \citenamefont {Aberle}}]{Ge2012}%
  \BibitemOpen
  \bibfield  {author} {\bibinfo {author} {\bibfnamefont {J.}~\bibnamefont
  {Ge}}, \bibinfo {author} {\bibfnamefont {Z.}~\bibnamefont {Ling}}, \bibinfo
  {author} {\bibfnamefont {J.}~\bibnamefont {Wong}}, \bibinfo {author}
  {\bibfnamefont {T.}~\bibnamefont {Mueller}},\ and\ \bibinfo {author}
  {\bibfnamefont {A.}~\bibnamefont {Aberle}},\ }\href
  {https://doi.org/https://doi.org/10.1016/j.egypro.2012.02.013} {\bibfield
  {journal} {\bibinfo  {journal} {Energy Procedia}\ }\textbf {\bibinfo {volume}
  {15}},\ \bibinfo {pages} {107 } (\bibinfo {year} {2012})}\BibitemShut
  {NoStop}%
\bibitem [{\citenamefont {Staebler}\ and\ \citenamefont
  {Wronski}(1977)}]{Staebler1977}%
  \BibitemOpen
  \bibfield  {author} {\bibinfo {author} {\bibfnamefont {D.~L.}\ \bibnamefont
  {Staebler}}\ and\ \bibinfo {author} {\bibfnamefont {C.~R.}\ \bibnamefont
  {Wronski}},\ }\href {https://doi.org/10.1063/1.89674} {\bibfield  {journal}
  {\bibinfo  {journal} {Applied Physics Letters}\ }\textbf {\bibinfo {volume}
  {31}},\ \bibinfo {pages} {292} (\bibinfo {year} {1977})}\BibitemShut
  {NoStop}%
\bibitem [{\citenamefont {Su}\ \emph {et~al.}(2002)\citenamefont {Su},
  \citenamefont {Taylor}, \citenamefont {Ganguly},\ and\ \citenamefont
  {Carlson}}]{Su2002}%
  \BibitemOpen
  \bibfield  {author} {\bibinfo {author} {\bibfnamefont {T.}~\bibnamefont
  {Su}}, \bibinfo {author} {\bibfnamefont {P.~C.}\ \bibnamefont {Taylor}},
  \bibinfo {author} {\bibfnamefont {G.}~\bibnamefont {Ganguly}},\ and\ \bibinfo
  {author} {\bibfnamefont {D.~E.}\ \bibnamefont {Carlson}},\ }\href
  {https://doi.org/10.1103/PhysRevLett.89.015502} {\bibfield  {journal}
  {\bibinfo  {journal} {Phys. Rev. Lett.}\ }\textbf {\bibinfo {volume} {89}},\
  \bibinfo {pages} {015502} (\bibinfo {year} {2002})}\BibitemShut {NoStop}%
\bibitem [{\citenamefont {Bobela}\ \emph {et~al.}(2011)\citenamefont {Bobela},
  \citenamefont {Branz},\ and\ \citenamefont {Stradins}}]{Bobela2011}%
  \BibitemOpen
  \bibfield  {author} {\bibinfo {author} {\bibfnamefont {D.~C.}\ \bibnamefont
  {Bobela}}, \bibinfo {author} {\bibfnamefont {H.~M.}\ \bibnamefont {Branz}},\
  and\ \bibinfo {author} {\bibfnamefont {P.}~\bibnamefont {Stradins}},\ }\href
  {https://doi.org/10.1063/1.3592264} {\bibfield  {journal} {\bibinfo
  {journal} {Applied Physics Letters}\ }\textbf {\bibinfo {volume} {98}},\
  \bibinfo {pages} {201908} (\bibinfo {year} {2011})}\BibitemShut {NoStop}%
\bibitem [{\citenamefont {Biswas}\ \emph {et~al.}(2007)\citenamefont {Biswas},
  \citenamefont {Atta-Fynn},\ and\ \citenamefont {Drabold}}]{Biswas2007}%
  \BibitemOpen
  \bibfield  {author} {\bibinfo {author} {\bibfnamefont {P.}~\bibnamefont
  {Biswas}}, \bibinfo {author} {\bibfnamefont {R.}~\bibnamefont {Atta-Fynn}},\
  and\ \bibinfo {author} {\bibfnamefont {D.~A.}\ \bibnamefont {Drabold}},\
  }\href {https://doi.org/10.1103/PhysRevB.76.125210} {\bibfield  {journal}
  {\bibinfo  {journal} {Phys. Rev. B}\ }\textbf {\bibinfo {volume} {76}},\
  \bibinfo {pages} {125210} (\bibinfo {year} {2007})}\BibitemShut {NoStop}%
\bibitem [{\citenamefont {Chakraborty}\ and\ \citenamefont
  {Drabold}(2009)}]{Chakraborty2009}%
  \BibitemOpen
  \bibfield  {author} {\bibinfo {author} {\bibfnamefont {S.}~\bibnamefont
  {Chakraborty}}\ and\ \bibinfo {author} {\bibfnamefont {D.~A.}\ \bibnamefont
  {Drabold}},\ }\href {https://doi.org/10.1103/PhysRevB.79.115214} {\bibfield
  {journal} {\bibinfo  {journal} {Phys. Rev. B}\ }\textbf {\bibinfo {volume}
  {79}},\ \bibinfo {pages} {115214} (\bibinfo {year} {2009})}\BibitemShut
  {NoStop}%
\bibitem [{\citenamefont {Biswas}\ and\ \citenamefont
  {Timilsina}(2011)}]{Biswas2011}%
  \BibitemOpen
  \bibfield  {author} {\bibinfo {author} {\bibfnamefont {P.}~\bibnamefont
  {Biswas}}\ and\ \bibinfo {author} {\bibfnamefont {R.}~\bibnamefont
  {Timilsina}},\ }\href {https://doi.org/10.1088/0953-8984/23/6/065801}
  {\bibfield  {journal} {\bibinfo  {journal} {J. Phys.: Cond. Matt.}\ }\textbf
  {\bibinfo {volume} {23}},\ \bibinfo {pages} {065801} (\bibinfo {year}
  {2011})}\BibitemShut {NoStop}%
\bibitem [{\citenamefont {Biswas}\ \emph {et~al.}(2017)\citenamefont {Biswas},
  \citenamefont {Paudel}, \citenamefont {Atta-Fynn}, \citenamefont {Drabold},\
  and\ \citenamefont {Elliott}}]{Biswas2017}%
  \BibitemOpen
  \bibfield  {author} {\bibinfo {author} {\bibfnamefont {P.}~\bibnamefont
  {Biswas}}, \bibinfo {author} {\bibfnamefont {D.}~\bibnamefont {Paudel}},
  \bibinfo {author} {\bibfnamefont {R.}~\bibnamefont {Atta-Fynn}}, \bibinfo
  {author} {\bibfnamefont {D.~A.}\ \bibnamefont {Drabold}},\ and\ \bibinfo
  {author} {\bibfnamefont {S.~R.}\ \bibnamefont {Elliott}},\ }\href
  {https://doi.org/10.1103/PhysRevApplied.7.024013} {\bibfield  {journal}
  {\bibinfo  {journal} {Phys. Rev. Applied}\ }\textbf {\bibinfo {volume} {7}},\
  \bibinfo {pages} {024013} (\bibinfo {year} {2017})}\BibitemShut {NoStop}%
\bibitem [{\citenamefont {Biswas}\ \emph {et~al.}(2020)\citenamefont {Biswas},
  \citenamefont {Paudel}, \citenamefont {Atta-Fynn},\ and\ \citenamefont
  {Elliott}}]{Biswas2020}%
  \BibitemOpen
  \bibfield  {author} {\bibinfo {author} {\bibfnamefont {P.}~\bibnamefont
  {Biswas}}, \bibinfo {author} {\bibfnamefont {D.}~\bibnamefont {Paudel}},
  \bibinfo {author} {\bibfnamefont {R.}~\bibnamefont {Atta-Fynn}},\ and\
  \bibinfo {author} {\bibfnamefont {S.~R.}\ \bibnamefont {Elliott}},\ }\href
  {https://doi.org/10.1039/C9NR08209C} {\bibfield  {journal} {\bibinfo
  {journal} {Nanoscale}\ }\textbf {\bibinfo {volume} {12}},\ \bibinfo {pages}
  {1464} (\bibinfo {year} {2020})}\BibitemShut {NoStop}%
\bibitem [{\citenamefont {Limbu}\ \emph {et~al.}(2020)\citenamefont {Limbu},
  \citenamefont {Elliott}, \citenamefont {Atta-Fynn},\ and\ \citenamefont
  {Biswas}}]{Limbu2020}%
  \BibitemOpen
  \bibfield  {author} {\bibinfo {author} {\bibfnamefont {D.~K.}\ \bibnamefont
  {Limbu}}, \bibinfo {author} {\bibfnamefont {S.~R.}\ \bibnamefont {Elliott}},
  \bibinfo {author} {\bibfnamefont {R.}~\bibnamefont {Atta-Fynn}},\ and\
  \bibinfo {author} {\bibfnamefont {P.}~\bibnamefont {Biswas}},\ }\href
  {https://doi.org/10.1038/s41598-020-64327-3} {\bibfield  {journal} {\bibinfo
  {journal} {Sci. Rep.}\ }\textbf {\bibinfo {volume} {10}},\ \bibinfo {pages}
  {7742} (\bibinfo {year} {2020})}\BibitemShut {NoStop}%
\bibitem [{\citenamefont {Biswas}\ \emph {et~al.}(2014)\citenamefont {Biswas},
  \citenamefont {Drabold},\ and\ \citenamefont {Atta-Fynn}}]{BiswasJAP2014}%
  \BibitemOpen
  \bibfield  {author} {\bibinfo {author} {\bibfnamefont {P.}~\bibnamefont
  {Biswas}}, \bibinfo {author} {\bibfnamefont {D.~A.}\ \bibnamefont
  {Drabold}},\ and\ \bibinfo {author} {\bibfnamefont {R.}~\bibnamefont
  {Atta-Fynn}},\ }\href {https://doi.org/10.1063/1.4905024} {\bibfield
  {journal} {\bibinfo  {journal} {Journal of Applied Physics}\ }\textbf
  {\bibinfo {volume} {116}},\ \bibinfo {pages} {244305} (\bibinfo {year}
  {2014})}\BibitemShut {NoStop}%
\bibitem [{\citenamefont {Guerrero}\ and\ \citenamefont
  {Strubbe}(2020)}]{Guerrero2020}%
  \BibitemOpen
  \bibfield  {author} {\bibinfo {author} {\bibfnamefont {E.}~\bibnamefont
  {Guerrero}}\ and\ \bibinfo {author} {\bibfnamefont {D.~A.}\ \bibnamefont
  {Strubbe}},\ }\href {https://doi.org/10.1103/PhysRevMaterials.4.025601}
  {\bibfield  {journal} {\bibinfo  {journal} {Phys. Rev. Materials}\ }\textbf
  {\bibinfo {volume} {4}},\ \bibinfo {pages} {025601} (\bibinfo {year}
  {2020})}\BibitemShut {NoStop}%
\bibitem [{\citenamefont {Sekimoto}\ \emph {et~al.}(2016)\citenamefont
  {Sekimoto}, \citenamefont {Matsumoto}, \citenamefont {Sagara}, \citenamefont
  {Hishida},\ and\ \citenamefont {Terakawa}}]{Sekimoto2016}%
  \BibitemOpen
  \bibfield  {author} {\bibinfo {author} {\bibfnamefont {T.}~\bibnamefont
  {Sekimoto}}, \bibinfo {author} {\bibfnamefont {M.}~\bibnamefont {Matsumoto}},
  \bibinfo {author} {\bibfnamefont {A.}~\bibnamefont {Sagara}}, \bibinfo
  {author} {\bibfnamefont {M.}~\bibnamefont {Hishida}},\ and\ \bibinfo {author}
  {\bibfnamefont {A.}~\bibnamefont {Terakawa}},\ }\href
  {https://doi.org/https://doi.org/10.1016/j.jnoncrysol.2016.05.030} {\bibfield
   {journal} {\bibinfo  {journal} {J. Non-Cryst. Solids}\ }\textbf {\bibinfo
  {volume} {447}},\ \bibinfo {pages} {207 } (\bibinfo {year}
  {2016})}\BibitemShut {NoStop}%
\bibitem [{\citenamefont {Sekimoto}\ \emph {et~al.}(2018)\citenamefont
  {Sekimoto}, \citenamefont {Matsumoto},\ and\ \citenamefont
  {Terakawa}}]{Sekimoto2018}%
  \BibitemOpen
  \bibfield  {author} {\bibinfo {author} {\bibfnamefont {T.}~\bibnamefont
  {Sekimoto}}, \bibinfo {author} {\bibfnamefont {M.}~\bibnamefont
  {Matsumoto}},\ and\ \bibinfo {author} {\bibfnamefont {A.}~\bibnamefont
  {Terakawa}},\ }\href {https://doi.org/10.7567/jjap.57.08rb07} {\bibfield
  {journal} {\bibinfo  {journal} {Jpn. J. Appl. Phys.}\ }\textbf {\bibinfo
  {volume} {57}},\ \bibinfo {pages} {08RB07} (\bibinfo {year}
  {2018})}\BibitemShut {NoStop}%
\bibitem [{\citenamefont {Paudel}\ \emph {et~al.}(2018)\citenamefont {Paudel},
  \citenamefont {Atta-Fynn}, \citenamefont {Drabold}, \citenamefont {Elliott},\
  and\ \citenamefont {Biswas}}]{Paudel2018}%
  \BibitemOpen
  \bibfield  {author} {\bibinfo {author} {\bibfnamefont {D.}~\bibnamefont
  {Paudel}}, \bibinfo {author} {\bibfnamefont {R.}~\bibnamefont {Atta-Fynn}},
  \bibinfo {author} {\bibfnamefont {D.~A.}\ \bibnamefont {Drabold}}, \bibinfo
  {author} {\bibfnamefont {S.~R.}\ \bibnamefont {Elliott}},\ and\ \bibinfo
  {author} {\bibfnamefont {P.}~\bibnamefont {Biswas}},\ }\href
  {https://doi.org/10.1103/PhysRevB.97.184202} {\bibfield  {journal} {\bibinfo
  {journal} {Phys. Rev. B}\ }\textbf {\bibinfo {volume} {97}},\ \bibinfo
  {pages} {184202} (\bibinfo {year} {2018})}\BibitemShut {NoStop}%
\bibitem [{\citenamefont {Wooten}\ \emph {et~al.}(1985)\citenamefont {Wooten},
  \citenamefont {Winer},\ and\ \citenamefont {Weaire}}]{W3}%
  \BibitemOpen
  \bibfield  {author} {\bibinfo {author} {\bibfnamefont {F.}~\bibnamefont
  {Wooten}}, \bibinfo {author} {\bibfnamefont {K.}~\bibnamefont {Winer}},\ and\
  \bibinfo {author} {\bibfnamefont {D.}~\bibnamefont {Weaire}},\ }\href
  {https://doi.org/10.1103/PhysRevLett.54.1392} {\bibfield  {journal} {\bibinfo
   {journal} {Phys. Rev. Lett.}\ }\textbf {\bibinfo {volume} {54}},\ \bibinfo
  {pages} {1392} (\bibinfo {year} {1985})}\BibitemShut {NoStop}%
\bibitem [{\citenamefont {Barkema}\ and\ \citenamefont
  {Mousseau}(2000)}]{Barkema2000}%
  \BibitemOpen
  \bibfield  {author} {\bibinfo {author} {\bibfnamefont {G.~T.}\ \bibnamefont
  {Barkema}}\ and\ \bibinfo {author} {\bibfnamefont {N.}~\bibnamefont
  {Mousseau}},\ }\href {https://doi.org/10.1103/PhysRevB.62.4985} {\bibfield
  {journal} {\bibinfo  {journal} {Phys. Rev. B}\ }\textbf {\bibinfo {volume}
  {62}},\ \bibinfo {pages} {4985} (\bibinfo {year} {2000})}\BibitemShut
  {NoStop}%
\bibitem [{\citenamefont {Soler}\ \emph {et~al.}(2002)\citenamefont {Soler},
  \citenamefont {Artacho}, \citenamefont {Gale}, \citenamefont {Garc{\'{\i}}a},
  \citenamefont {Junquera}, \citenamefont {Ordej{\'{o}}n},\ and\ \citenamefont
  {S{\'{a}}nchez-Portal}}]{Siesta2002}%
  \BibitemOpen
  \bibfield  {author} {\bibinfo {author} {\bibfnamefont {J.~M.}\ \bibnamefont
  {Soler}}, \bibinfo {author} {\bibfnamefont {E.}~\bibnamefont {Artacho}},
  \bibinfo {author} {\bibfnamefont {J.~D.}\ \bibnamefont {Gale}}, \bibinfo
  {author} {\bibfnamefont {A.}~\bibnamefont {Garc{\'{\i}}a}}, \bibinfo {author}
  {\bibfnamefont {J.}~\bibnamefont {Junquera}}, \bibinfo {author}
  {\bibfnamefont {P.}~\bibnamefont {Ordej{\'{o}}n}},\ and\ \bibinfo {author}
  {\bibfnamefont {D.}~\bibnamefont {S{\'{a}}nchez-Portal}},\ }\href
  {https://doi.org/10.1088\%2F0953-8984\%2F14\%2F11\%2F302} {\bibfield
  {journal} {\bibinfo  {journal} {J. Phys.: Condens. Matter}\ }\textbf
  {\bibinfo {volume} {14}},\ \bibinfo {pages} {2745} (\bibinfo {year}
  {2002})}\BibitemShut {NoStop}%
\bibitem [{\citenamefont {Mahan}\ \emph {et~al.}(2001)\citenamefont {Mahan},
  \citenamefont {Xu}, \citenamefont {Williamson}, \citenamefont {Beyer},
  \citenamefont {Perkins}, \citenamefont {Vanecek}, \citenamefont {Gedvilas},\
  and\ \citenamefont {Nelson}}]{Mahan2001}%
  \BibitemOpen
  \bibfield  {author} {\bibinfo {author} {\bibfnamefont {A.~H.}\ \bibnamefont
  {Mahan}}, \bibinfo {author} {\bibfnamefont {Y.}~\bibnamefont {Xu}}, \bibinfo
  {author} {\bibfnamefont {D.~L.}\ \bibnamefont {Williamson}}, \bibinfo
  {author} {\bibfnamefont {W.}~\bibnamefont {Beyer}}, \bibinfo {author}
  {\bibfnamefont {J.~D.}\ \bibnamefont {Perkins}}, \bibinfo {author}
  {\bibfnamefont {M.}~\bibnamefont {Vanecek}}, \bibinfo {author} {\bibfnamefont
  {L.~M.}\ \bibnamefont {Gedvilas}},\ and\ \bibinfo {author} {\bibfnamefont
  {B.~P.}\ \bibnamefont {Nelson}},\ }\href {https://doi.org/10.1063/1.1407317}
  {\bibfield  {journal} {\bibinfo  {journal} {Journal of Applied Physics}\
  }\textbf {\bibinfo {volume} {90}},\ \bibinfo {pages} {5038} (\bibinfo {year}
  {2001})}\BibitemShut {NoStop}%
\bibitem [{\citenamefont {Young}\ \emph {et~al.}(2007)\citenamefont {Young},
  \citenamefont {Stradins}, \citenamefont {Xu}, \citenamefont {Gedvilas},
  \citenamefont {Iwaniczko}, \citenamefont {Yan}, \citenamefont {Branz},
  \citenamefont {Wang.},\ and\ \citenamefont {Williamson}}]{Young2007}%
  \BibitemOpen
  \bibfield  {author} {\bibinfo {author} {\bibfnamefont {D.~L.}\ \bibnamefont
  {Young}}, \bibinfo {author} {\bibfnamefont {P.}~\bibnamefont {Stradins}},
  \bibinfo {author} {\bibfnamefont {Y.}~\bibnamefont {Xu}}, \bibinfo {author}
  {\bibfnamefont {L.~M.}\ \bibnamefont {Gedvilas}}, \bibinfo {author}
  {\bibfnamefont {E.}~\bibnamefont {Iwaniczko}}, \bibinfo {author}
  {\bibfnamefont {Y.}~\bibnamefont {Yan}}, \bibinfo {author} {\bibfnamefont
  {H.~M.}\ \bibnamefont {Branz}}, \bibinfo {author} {\bibfnamefont
  {Q.}~\bibnamefont {Wang.}},\ and\ \bibinfo {author} {\bibfnamefont {D.~L.}\
  \bibnamefont {Williamson}},\ }\href {https://doi.org/10.1063/1.2435959}
  {\bibfield  {journal} {\bibinfo  {journal} {Applied Physics Letters}\
  }\textbf {\bibinfo {volume} {90}},\ \bibinfo {pages} {081923} (\bibinfo
  {year} {2007})}\BibitemShut {NoStop}%
\bibitem [{voi()}]{void}%
  \BibitemOpen
  \href@noop {} {}\bibinfo {note} {The effective width of the void surface is
  chosen to be about 3 {\AA}. This is based on the results from the test
  calculations at 700 K, which show that the majority of silicon atoms near the
  void surface can move up to a distance of 2.5--3.0 {\AA}. This distance is
  roughly equal to the first nearest-neighbor distance between Si atoms in the
  amorphous environment. For structural analysis of voids, we thus use a
  void-surface region of 8 {\AA}.}\BibitemShut {Stop}%
\bibitem [{\citenamefont {Junquera}\ \emph {et~al.}(2001)\citenamefont
  {Junquera}, \citenamefont {Paz}, \citenamefont {S\'anchez-Portal},\ and\
  \citenamefont {Artacho}}]{Artacho2001}%
  \BibitemOpen
  \bibfield  {author} {\bibinfo {author} {\bibfnamefont {J.}~\bibnamefont
  {Junquera}}, \bibinfo {author} {\bibfnamefont {O.}~\bibnamefont {Paz}},
  \bibinfo {author} {\bibfnamefont {D.}~\bibnamefont {S\'anchez-Portal}},\ and\
  \bibinfo {author} {\bibfnamefont {E.}~\bibnamefont {Artacho}},\ }\href
  {https://doi.org/10.1103/PhysRevB.64.235111} {\bibfield  {journal} {\bibinfo
  {journal} {Phys. Rev. B}\ }\textbf {\bibinfo {volume} {64}},\ \bibinfo
  {pages} {235111} (\bibinfo {year} {2001})}\BibitemShut {NoStop}%
\bibitem [{\citenamefont {Troullier}\ and\ \citenamefont
  {Martins}(1991)}]{TM1991}%
  \BibitemOpen
  \bibfield  {author} {\bibinfo {author} {\bibfnamefont {N.}~\bibnamefont
  {Troullier}}\ and\ \bibinfo {author} {\bibfnamefont {J.}~\bibnamefont
  {Martins}},\ }\href {https://doi.org/10.1103/PhysRevB.43.1993} {\bibfield
  {journal} {\bibinfo  {journal} {Phys. Rev. B}\ }\textbf {\bibinfo {volume}
  {43}},\ \bibinfo {pages} {1993} (\bibinfo {year} {1991})}\BibitemShut
  {NoStop}%
\bibitem [{\citenamefont {Perdew}\ and\ \citenamefont {Zunger}(1981)}]{LDA}%
  \BibitemOpen
  \bibfield  {author} {\bibinfo {author} {\bibfnamefont {J.~P.}\ \bibnamefont
  {Perdew}}\ and\ \bibinfo {author} {\bibfnamefont {A.}~\bibnamefont
  {Zunger}},\ }\href {https://doi.org/10.1103/PhysRevB.23.5048} {\bibfield
  {journal} {\bibinfo  {journal} {Phys. Rev. B}\ }\textbf {\bibinfo {volume}
  {23}},\ \bibinfo {pages} {5048} (\bibinfo {year} {1981})}\BibitemShut
  {NoStop}%
\bibitem [{\citenamefont {Perdew}\ \emph {et~al.}(1996)\citenamefont {Perdew},
  \citenamefont {Burke},\ and\ \citenamefont {Ernzerhof}}]{PBE}%
  \BibitemOpen
  \bibfield  {author} {\bibinfo {author} {\bibfnamefont {J.~P.}\ \bibnamefont
  {Perdew}}, \bibinfo {author} {\bibfnamefont {K.}~\bibnamefont {Burke}},\ and\
  \bibinfo {author} {\bibfnamefont {M.}~\bibnamefont {Ernzerhof}},\ }\href
  {https://doi.org/10.1103/PhysRevLett.77.3865} {\bibfield  {journal} {\bibinfo
   {journal} {Phys. Rev. Lett.}\ }\textbf {\bibinfo {volume} {77}},\ \bibinfo
  {pages} {3865} (\bibinfo {year} {1996})}\BibitemShut {NoStop}%
\bibitem [{\citenamefont {Nos\'e}(1984)}]{Nose1984}%
  \BibitemOpen
  \bibfield  {author} {\bibinfo {author} {\bibfnamefont {S.}~\bibnamefont
  {Nos\'e}},\ }\href {https://doi.org/10.1063/1.447334} {\bibfield  {journal}
  {\bibinfo  {journal} {The Journal of Chemical Physics}\ }\textbf {\bibinfo
  {volume} {81}},\ \bibinfo {pages} {511} (\bibinfo {year} {1984})}\BibitemShut
  {NoStop}%
\bibitem [{\citenamefont {Kageyama}\ \emph {et~al.}(2011)\citenamefont
  {Kageyama}, \citenamefont {Akagawa},\ and\ \citenamefont
  {Fujiwara}}]{Kageyama2011}%
  \BibitemOpen
  \bibfield  {author} {\bibinfo {author} {\bibfnamefont {S.}~\bibnamefont
  {Kageyama}}, \bibinfo {author} {\bibfnamefont {M.}~\bibnamefont {Akagawa}},\
  and\ \bibinfo {author} {\bibfnamefont {H.}~\bibnamefont {Fujiwara}},\ }\href
  {https://doi.org/10.1103/PhysRevB.83.195205} {\bibfield  {journal} {\bibinfo
  {journal} {Phys. Rev. B}\ }\textbf {\bibinfo {volume} {83}},\ \bibinfo
  {pages} {195205} (\bibinfo {year} {2011})}\BibitemShut {NoStop}%
\bibitem [{\citenamefont {Beyer}(2003)}]{Beyer2003}%
  \BibitemOpen
  \bibfield  {author} {\bibinfo {author} {\bibfnamefont {W.}~\bibnamefont
  {Beyer}},\ }\href {https://doi.org/10.1016/S0927-0248(02)00438-5} {\bibfield
  {journal} {\bibinfo  {journal} {Solar Energy Materials and Solar Cells}\
  }\textbf {\bibinfo {volume} {78}},\ \bibinfo {pages} {235 } (\bibinfo {year}
  {2003})}\BibitemShut {NoStop}%
\bibitem [{\citenamefont {Beyer}\ and\ \citenamefont
  {Wagner}(1982)}]{Beyer1982}%
  \BibitemOpen
  \bibfield  {author} {\bibinfo {author} {\bibfnamefont {W.}~\bibnamefont
  {Beyer}}\ and\ \bibinfo {author} {\bibfnamefont {H.}~\bibnamefont {Wagner}},\
  }\href {https://doi.org/10.1063/1.330474} {\bibfield  {journal} {\bibinfo
  {journal} {Journal of Applied Physics}\ }\textbf {\bibinfo {volume} {53}},\
  \bibinfo {pages} {8745} (\bibinfo {year} {1982})}\BibitemShut {NoStop}%
\bibitem [{\citenamefont {Sekimoto}\ \emph {et~al.}(2014)\citenamefont
  {Sekimoto}, \citenamefont {Matsumoto}, \citenamefont {Hishida},\ and\
  \citenamefont {Terakawa}}]{Sekimoto2014}%
  \BibitemOpen
  \bibfield  {author} {\bibinfo {author} {\bibfnamefont {T.}~\bibnamefont
  {Sekimoto}}, \bibinfo {author} {\bibfnamefont {M.}~\bibnamefont {Matsumoto}},
  \bibinfo {author} {\bibfnamefont {M.}~\bibnamefont {Hishida}},\ and\ \bibinfo
  {author} {\bibfnamefont {A.}~\bibnamefont {Terakawa}},\ }\href
  {https://doi.org/10.7567/JJAP.53.095501} {\bibfield  {journal} {\bibinfo
  {journal} {Jpn. J. Appl. Phys.}\ }\textbf {\bibinfo {volume} {53}},\ \bibinfo
  {pages} {095501} (\bibinfo {year} {2014})}\BibitemShut {NoStop}%
\bibitem [{\citenamefont {Chabal}\ and\ \citenamefont
  {Patel}(1984)}]{Chabal1984}%
  \BibitemOpen
  \bibfield  {author} {\bibinfo {author} {\bibfnamefont {Y.~J.}\ \bibnamefont
  {Chabal}}\ and\ \bibinfo {author} {\bibfnamefont {C.~K.~N.}\ \bibnamefont
  {Patel}},\ }\href {https://doi.org/10.1103/PhysRevLett.53.210} {\bibfield
  {journal} {\bibinfo  {journal} {Phys. Rev. Lett.}\ }\textbf {\bibinfo
  {volume} {53}},\ \bibinfo {pages} {210} (\bibinfo {year} {1984})}\BibitemShut
  {NoStop}%
\bibitem [{\citenamefont {Biswas}\ and\ \citenamefont
  {Elliott}(2015)}]{BiswasIOP2015}%
  \BibitemOpen
  \bibfield  {author} {\bibinfo {author} {\bibfnamefont {P.}~\bibnamefont
  {Biswas}}\ and\ \bibinfo {author} {\bibfnamefont {S.~R.}\ \bibnamefont
  {Elliott}},\ }\href@noop {} {\bibfield  {journal} {\bibinfo  {journal} {J.
  Phys.: Cond. Matt.}\ }\textbf {\bibinfo {volume} {27}},\ \bibinfo {pages}
  {435201} (\bibinfo {year} {2015})}\BibitemShut {NoStop}%
\bibitem [{\citenamefont {Kokalj}(1999)}]{xc}%
  \BibitemOpen
  \bibfield  {author} {\bibinfo {author} {\bibfnamefont {A.}~\bibnamefont
  {Kokalj}},\ }\href {https://doi.org/10.1016/S1093-3263(99)00028-5} {\bibfield
   {journal} {\bibinfo  {journal} {J. Mol. Graph. Model.}\ }\textbf {\bibinfo
  {volume} {17}},\ \bibinfo {pages} {176 } (\bibinfo {year}
  {1999})}\BibitemShut {NoStop}%
\bibitem [{\citenamefont {Wadell}(1935)}]{Wadell1935}%
  \BibitemOpen
  \bibfield  {author} {\bibinfo {author} {\bibfnamefont {H.}~\bibnamefont
  {Wadell}},\ }\href {https://doi.org/10.1086/624298} {\bibfield  {journal}
  {\bibinfo  {journal} {J. Geology}\ }\textbf {\bibinfo {volume} {43}},\
  \bibinfo {pages} {250} (\bibinfo {year} {1935})}\BibitemShut {NoStop}%
\end{thebibliography}

\end{document}